\documentclass[10pt]{article}
\usepackage{graphicx}
\usepackage{authblk}

\usepackage{natbib}
\usepackage{amssymb,amsthm,amsmath}
\usepackage{xcolor,paralist,hyperref,titlesec,fancyhdr,etoolbox}
\usepackage[utf8]{inputenc}

\hypersetup{ colorlinks=false, linkcolor=black, filecolor=black, urlcolor=black }

\usepackage{lipsum}

\DeclareMathOperator*{\argmin}{arg\,min}

\begin{document}
\title{The Typicality of Regimes Associated with Northern Hemisphere Heatwaves} %%%%%%%%%%%%
\author{Christopher~C.~Chapman\textsuperscript{a}, Didier~P.~Monselesan\textsuperscript{a}, James~S.~Risbey\textsuperscript{a}, Abdelwaheb~Hannachi\textsuperscript{b}, Valerio Lucarini\textsuperscript{c} and Richard~Matear\textsuperscript{a}}

\affil[a]{CSIRO Environment, Hobart, Tasmania, Australia}
\affil[b]{Department of Meteorology and Bolin Centre for Climate Research, Stockholm University, Stockholm, Sweden}
\affil[c]{School of Computing and Mathematical Sciences, University of Leicester, Leicester, UK}

\date{\today}
%\address{Address}
%\email{example@mail.com}
\maketitle

\let\thefootnote\relax
\footnotetext{Submitted to the Journal of Climate} %%%%%%%%%%

\begin{abstract}
We study the hemispheric to continental scale regimes that lead to summertime heatwaves in the Northern Hemisphere. By using a powerful data mining methodology -- archetype analysis -- we identify characteristic spatial patterns consisting of a blocking high pressure systems embedded within a meandering upper atmosphere circulation that is longitudinally modulated by coherent Rossby Wave Packets. Periods when these atmospheric regimes are strongly expressed correspond to large increases in the likelihood of extreme surface temperature. Most strikingly, these regimes are shown to be typical of surface extremes and frequently reoccur. Three well publicised heatwaves are studied in detail - the June-July 2003 western European heatwave, the August 2010 ``Russian" heatwave, and the June 2021 ``Heatdome" event across western North America, and are shown to be driven by blocking high pressure systems linked to stalled Rossby Wave Packets. We discuss the implications of our work for long-range prediction or early warning, climate model assessment and post-event diagnosis.
\end{abstract} %%%%%%%%%

\bigskip

Heatwaves are among the deadliest and most costly natural hazards, with devastating effects on human societies, ecosystems, agriculture and infrastructure \citep{Poumadere2005,Perkins2015, BarriopedroEtAl2011, BarriopedroEtAl2023,ParkEtAl2023}. Climate change is expected to increase both the intensity, frequency, and persistence of these events, with more regions of the globe likely to experience extreme heat, more frequently \citep{StottEtAl2005,PerkinsEtAl2012, DomeisenEtAl2022}. \newline

At the local or regional scale, heatwaves are the result of a complex interplay between the large-scale atmospheric circulation, thermodynamics and local-scale processes such as land use and soil moisture \citep{CoumouEtAl2014, Perkins2015, HauserEtAl2016, BartusekEtAl2022}. However, in the mid-latitudes and sub-tropics, certain atmospheric conditions are common to most, if not all, extreme temperature events, chief among them blocking high pressure systems. These persistent features direct anomalous airflow over a region, and enhance surface heating through shortwave radiative heating (blocks are often associated with clear sky conditions), and adiabatic heating by subsidence of air parcels  \citep{Ghil2002,WoollingsEtAl2018, Lupo2021}. There is strong evidence that persistent blocks are a fundamental emergent feature of the chaotic mid-latitude atmospheric circulation \citep{Charney1979, CharneyStraus1980,Benzi1986,Malguzzi1997,Schubert2016}. \newline

 Blocking systems are themselves commonly embedded within a broader wavy atmospheric circulation  \citep{KornhuberEtAl2017}. On occasion, these meandering circulations can span an entire hemisphere, leading  long distance teleconnections \citep{bra02:jcli, PetoukhovEtAl2013, KornhuberEtAl2019} that can result in concurrent extremes in regions separated by large distances. Examples include simultaneous heatwaves in Europe and eastern Asia \citep{Lau&Kim2012, CoumouEtAl2014, DengEtAl2018, KornhuberEtAl2019, HaoEtAl2022} or, in the much studied case of the 2010 ``Russian" heatwave \citep{Dole2011} the co-occurrence of catastrophic floods in Northern Pakistan \citep{Hong2011,Lau&Kim2012,Galfi2021PRL}. Regional modulation of the amplitude of the large scale circulation occurs due to spatially confined envelopes of greater or lesser amplitudes of the perturbed flow, that are known as Rossby Wave Packets (herin RWPs \citet{WirthEtAl2018,FragkoulidisEtAl2018}). RWPs propagate at the group velocity, which is typically slower than the phase velocity of the individual peaks and troughs contained within the envelope \citep{Lee&Held1993, WirthEtAl2018}. They can remain coherent for long periods of time, despite the chaotic evolution of the systems within it, and are shaped by their interaction with the background flow and orography. Exploiting the slow evolution of RWPs has been identified as potential a route for extended range prediction of extreme events \citep{TengEtAL2013,Grazzini&Vitart2015,Pyrina&Domesien2023,WhiteEtAl2022,JimenezEstev&Domeisen2022} \newline
 
  %, although lower wavenumber configurations (corresponding to larger spatial scales) have also been implicated in extreme events \citep{YangEtAl2024,DengEtAl2018}.  As a consequence of their impact on extreme events and their persistent coherence, Rossby wave packets have been potential target for extended range prediction systems \citep{Grazzini&Vitart2015, Quinting&Vitart2019}.

%Research into links between continental scale patterns and surface heatwaves has been extremely active for several decades. From a climate perspective, understanding how the continental scale atmospheric circulation effects extremes is important for assessing projections of future heatwaves, particularly those derived from coarse resolution climate models incapable of correctly representing regional effects \citep{Perkins2015,MannEtAl2017,MannEtAl2018,DomeisenEtAl2022}. While a climate model may not be able to accurately simulation regional processes driven by small scale phenomena such as clouds, topography, and soil and vegetation feedbacks, it \textit{may} be capable of simulating the broad-scale patterns that increase the probability of heatwaves over certain regions. Secondly, from a forecasting perspective, predictability of heatwaves beyond the usual 5-7 days provided by standard numerical weather forecasts, remains limited \citep{DomeisenEtAl2022}. In addition, while the current generation of extended range forecasts (sub-seasonal to seasonal scale) provide some skill, there remains a significant skill gap from 1 week to two months. . \newline

Using methods taken from statistics and dynamical systems, recent studies of the climate extremes have shown that, somewhat against intuition, certain extreme states are \textit{typical}. Using a statistical approach known as large deviation theory, \citet{Galfi2021PRL, Galfi2021,LucariniEtAl2023} and \citet{NoyelleEtAl2023} showed that the continental or larger scale circulation anomalies at a single location reassemble each other when the underlying statistics are conditioned on large deviations (i.e. extremes) from the climatological state. The use of rare-event algorithms \citep{Ragone2018,Ragone2021} and deep learning \citep{JacquesDumas2022} makes it possible to efficiently extract such patterns and better interpret the associated physical processes. Remarkably, as one considers more extreme events, their associated atmospheric configurations cluster more and more closely, which implies that the extreme states of the atmosphere are indeed typical, albeit rare, states in the conditioned statistics \citep{Galfi2021,NoyelleEtAl2023}. \newline

Independently \citet{RisbeyEtAl2023} and \citet{Fisher2023} demonstrated in studies of the 2021 Western North America ``heatdome" heatwave that a particular atmospheric state, consisting of a blocking high embedded in a larger-scale wavetrain, was required to generate extreme temperatures on the west coast of North America. \citet{RisbeyEtAl2023}, using a large climate model ensemble under present day conditions, noted that only a ``handful of summer days among the millions simulated have strong pattern match with the hottest model day". \citet{Fisher2023} demonstrated using a boosted ensemble that large-scale circulation associated with extreme events showed strong dynamic similarity across a large number of realisations of the current climate state. Notably, the large-scale patterns were robust to perturbations imposed on the atmospheric state, although the intensity of simulated events was sensitive to those perturbations -- resulting in temperatures that were occasionally larger than those ever observed in the historical record. These ensemble based studies validate the notion of dynamical typicality for heatwave events proposed in \cite{LucariniEtAl2023}. \newline   

With notable exceptions (e.g.~\citet{Springeretal2024}), almost all previous work into both continental-scale drivers of extreme conditions and their typicality are based on statistics at either one or more geographical locations. From these point statistics, most studies then ``zoom out" and seek to link the larger-scale atmospheric circulation to the local extreme, described by \citet{cha22:ncom} as an \textit{inside-out} approach. However, as noted by \citet{RisbeyEtAl2023} and \citet{Fisher2023}, the temperature at a fixed location is sensitive to the precise positioning and evolution of the responsible weather systems. Similarly, the chaotic nature of the atmosphere means that no large-scale pattern, even one associated with extremes, is ever likely to exactly repeat \citep{Lorenz1969}. Additionally, local and regional-scale processes, such as the non-linear interactions between the atmosphere and land surface, also strongly influence temperatures at the local-scale \citep{BartusekEtAl2022,Conrick&Mass2023}, which further complicates comparisons between events. \newline

%As a result, identifying two or more atmospheric regimes with similar effects at a single geographical location is subject to some ambiguity and inevitable subjectivity. We remark that these issues are eased when variables are averaged over temporal and spatial scales larger than those associated with usual synoptic variability, and that, while truly good analogues can hardly be found in practice, large deviation laws control, via the so-called concentration property, the quality of the analogues in the limit where  \textit{very} extreme events are selected \citep{Galfi2021,Galfi2021PRL}. \newline

Here, we look at surface temperature extremes from a global, physically based angle. In this paper, we take an \textit{outside-in} approach, by directly and unambiguously identifying the continental to hemispheric scale patterns associated with regional summer time surface heatwaves in the Northern Hemisphere. To do so, we use a powerful data-driven method called \textit{Archetype Analysis} (AA) that robustly selects global extreme configurations from high-dimensional datasets in an unsupervised manner. AA has been applied in studies of extreme precipitation \citep{ste15:jcli}, persistent atmospheric blocking \citep{ris21:mwr}, marine heatwaves \citep{cha22:ncom}, and El-Ni\~{n}o characterisation \citep{mon24:arxiv}.  Once the large-scale extreme regimes have been identified, we are then able to zoom-in to investigate the regional and local impacts. \newline

Using AA, we extract the large-scale patterns associated with extreme surface temperatures in the extratropical Northern Hemisphere. Well-studied events, such as the 2003 western European heatwave \citep{BlackEtAl2004,DuchezEtAl2016}, the 2010 Eastern European or ``Russian" heatwave \citep{Dole2011,BarriopedroEtAl2011,Lau&Kim2012,HauserEtAl2016}, and the 2021 western North-America ``heatdome" event \citep{BartusekEtAl2022, WhiteEtAl2023}, appear naturally from our analysis, as well as several less-well studied events. We will provide a plausible dynamical mechanism by linking the extreme events with hemispheric-scale wave patterns and continental-scale RWPs. Although we focus on the Northern Hemisphere, similar large-scale circulation anomalies have been observed in the Southern Hemisphere (e.g. \citet{ParkerEtAl2014}) and our results should have broad applicability.\newline  

The remainder of this paper is organised as follows: Section \ref{Sec:Methods_and_Data} describes our methodology and data sources, including a brief discussion of AA. In section \ref{Sec:Results} we evaluate the approach using both broad scale patterns and event based investigation. Section \ref{Sec:Dynamics} provides dynamical insight into the large scale extreme patterns, and we conclude in  Section \ref{Sec:Discussion_Conclusion} with a discussion of the implications arising from this work, its limitations, and several avenues for future work.

\section{Data and Methods}  \label{Sec:Methods_and_Data}

\subsection{Data}

\subsubsection{Atmospheric Reanalysis}

The primary data source used here is the output from the  Japanese 55-Year Reanalysis project (JRA-55, \cite{kob15:jmsj}), with the extended output from January 1st 1958 until the 31st of December 2023, a total of 65 years. We make use of the daily mean surface air temperature (at 2m above the ground or sea-surface), as well as the 500hPa geopotential height and 200-hPa wind velocities. \newline

We restrict our attention to an extended Northern Hemisphere (boreal) summer, from the 1st of May until the 30th of September. We have made this choice to concentrate on events with the highest absolute temperatures, as these events that typically have the greatest direct impact on society. However, we note that heatwave conditions during cooler seasons can also have strong impacts on ecology and agriculture. We use anomalies relative to a day-of-year climatology computed from the full data period. To remove the influence on ongoing climate change, we have subtracted the linear trend at every grid point, although we note that a residual higher order trends may still exist. \newline 

We restrict our attention to the Northern Hemisphere by including only latitudes between the equator and 90$^{\circ}$N. We focus on the Northern Hemisphere as the larger human population means that extreme heatwave events are generally better observed and more studied than Southern Hemisphere equivalents and the body of literature is more extensive. However, the Southern Hemisphere is not immune to heatwave conditions \citep{ParkerEtAl2014,par14:grl,QuintingEtAl2018} and extending our methodology south of the equator will be the subject of continuing work.

\subsection{Methods} \label{Sec:Methods}

\subsubsection{Archetype Analysis}

Archetype Analysis \citep{cut94:tec, han17:jcli} (herein AA), is a matrix factorisation method that can extract extreme or outlying configurations from a dataset of finite but otherwise arbitrary dimension. Consider a data matrix $\mathbf{X} = \mathbf{X}_{s \times t}$ describing a spatial-temporal field where $s$ corresponds to the set of spatial dimensions (e.g. latitude and longitude) and $t$ is time. AA approximates $\mathbf{X}$ as the convex combination of a set of $p$ archetypal spatial patterns $\mathbf{Z}_{s \times p}$,
\begin{equation} \label{Eqn:AA_1}
\mathbf{X}_{s \times t} \approx  \mathbf{Z}_{s \times p} \mathbf{S}_{p \times t}.
\end{equation}
where $p<t$ (and typically $p\ll t$), resulting in a reduction in the dimension of the dataset. 

The elements $S_{ij}$ of $\mathbf{S}_{p \times t}$, $i=1,\ldots,p$ for all times $j=1,\ldots,t$, with $S_{ij} \geq 0$, are the convex weights that we term the affiliation probability, which is subject to the convexity constraints $\sum^{p}_{i=1} S_{it}=1, \forall t$. The archetypal patterns are themselves constructed from a (generally sparse) convex combination of individual snapshots from the data:
\begin{equation} \label{Eqn:AA_2}
\mathbf{Z}_{s \times p} = \mathbf{X}_{s \times t} \mathbf{C}_{t \times p},
\end{equation}   
where the matrix $\mathbf{C}_{t \times p}$ is subject to the constraints $\sum^{t}_{i=1} C_{ip}=1$ with $C_{ip} \geq 0, \forall p$. 

%Combining  gives
%\begin{linenomath*}
%\begin{equation}\label{Eqn:AA_3}
%\mathbf{X}_{s \times t} \approx \mathbf {X}_{s \times %t}\mathbf{C}_{t \times p} \mathbf{S}_{p \times t}.
%\end{equation}
%\end{linenomath*}

For a given data matrix $\mathbf{X}$, the challenge is to determine the stochastic weight matrices $\mathbf{S}$ and $\mathbf{C}$. To do so, we combine Eqns. \ref{Eqn:AA_1} and \ref{Eqn:AA_2} and solve the resulting non-linear optimisation routine:
%\begin{linenomath*}
\begin{equation} \label{Eqn:AA_min}
%   \{ \mathbf{S},\mathbf{C} \}  = \argmin_{\mathbf{S},\mathbf{C}} \| \mathbf{X} - \mathbf{XCS} \|_{F}
   \argmin_{\mathbf{S},\mathbf{C}} \| \mathbf{X} - \mathbf{XCS} \|_{F}
\end{equation}
%\end{linenomath*}
where $\|\cdot\|_{F}$ denotes the Froebenius norm, which is simply the sum of squares of the absolute value of all matrix elements. \newline

We obtain numerical approximations to Eqn. \ref{Eqn:AA_min} using the reduced space approach described in \cite{bla22:aies} and \cite{cha22:ncom}. We first reduce the dimension of the dataset using principle component analysis, retaining 95\% of the variance and weight the data-matrix by the square root of the cosine of latitude. The optimization problem in Eqn. \ref{Eqn:AA_min} is solved 1,000 times with different initialisation using an efficient projection-gradient method, in order to avoid inadvertently selecting a local minimum solution and provide broad coverage of the solution space. The solution with the smallest relative sum squared error (see Eqn. 4 of \citet{bla22:aies} ) is taken as the optimal solution. \newline

\subsubsection{Interpreting the Output of Archetype Analysis}

At time of writing, AA is not yet in widespread use in geophysics. To aid the reader, we provide a brief explanation  of the utility of AA and its interpretation. \newline

AA approximates the data as a number $p<<t$ of spatial patterns $\mathbf{Z}$, each with an associated affiliation probability timeseries $\mathbf{S}_{p \times t}$. At every time step, the affiliation probability gives the probability that the spatial pattern associated with archetype $k \in p$ is expressed. As such, when $S_{kt} \rightarrow 1$, the data for that snapshot will resemble the archetypal spatial pattern. Conversely, at times when $S_{kt}\rightarrow 0$, we can be reasonably certain that the data will not resemble the $k$th archetype. \newline

The spatial patterns themselves are constructed by a convex weighted average of all snapshots in the dataset. In practice, these weights (the $\mathbf{C}$-matrix in Eqn. \ref{Eqn:AA_2}) are zero for almost every snapshot, so that only a few very special snapshots contribute to the construction of an archetype. $\mathbf{C}$ is associated with the learning phase of the algorithm, whilst $\mathbf{S}$ encodes the reduced-order description of the atmospheric state. It can be shown that, due to the convexity constraints applied to the optimisation problem, those snapshots with non-zero $\mathbf{C}$-matrix weights should lie close to the convex hull, or bounding envelope, of the dataset \citep{cut94:tec}. AA thus provides a discrete approximation to the convex hull by selecting extremes of the observations in the high-dimensional space. As it has been shown that in a chaotic dynamical system the extremes of well-behaved physical observables are found at the boundaries of the attractor \citep{Lucarinietal2014JSP}, we take the states identified by AA as the extreme configurations of the system.\newline

The number of archetypes, $p$, is chosen by the user and is typically much smaller than the number of time-snapshots in the dataset. However, experience has shown that with only a small number of archetypes, one can represent a surprisingly large variety of extreme events. Previous work \citep{han17:jcli, ric21:jhm, ris21:mwr, bla22:aies} has demonstrated that the choice of the number of archetypes is critical: too few and we lose the ability to discriminate between dissimilar events, too many and we lose the simplification of the dataset into a conceptually tractable number of states. This issue is discussed in more detail in Appendix A. It is also important to note that the state at any one time step may be best approximated as a mixture of two or more archetypes. 

\subsubsection{Weighted Composites from Archetypal Affiliations}

As noted in \cite{bla22:aies} and \cite{cha22:ncom}, the affiliation time series for a given archetype $p$ can be used to generate composites of any ancillary variable with data matrix $\mathbf{Y}$, provided that there is an overlapping time period of that variable and the variable that was used to calculate the archetypes. For example, in this work we calculate the average 500 hPa geopotential height patterns that occur with the surface temperature patterns obtained from AA. Affiliation composites, that is using the $\mathbf{S}$-matrix time series, for all $p$ archetypes can be computed by a simple weighted average:
\begin{equation} \label{Eqn:S_composite}
\overline{Y}_{sp} = \frac{\sum^{t}_{i} Y_{si}S_{pi}} {\sum^{t}_{i} S_{pi} },
\end{equation}   
whereas composites formed using the $\mathbf{C}$-matrix weights are computed as:
\begin{equation} \label{Eqn:C_composite}
\overline{Y}_{cp} = \sum^{t}_{i} Y_{ci}C_{pi}.
\end{equation}   
since the convexity constraints impose $\sum^{t}_{i} C_{pi}=1$. In practice, the spatial patterns produced by $\mathbf{C}$-matrix or $\mathbf{S}$-matrix composites as similar. However, the patterns from $\mathbf{C}$-matrix composites tend to have larger magnitudes as the weights select for points closer to the edge of the distribution, while the $\mathbf{S}$-matrix weights fall closer to the center. Additionally, the sparsity of the $\mathbf{C}$-matrix weights can result in noisy composites or those without statistical power, while $\mathbf{S}$-matrix composites are generally smoother as they are constructed from a larger sample population.

\subsubsection{Event Identification and Regime Determination} \label{Sec:Event_Identification}

We use the AA to determine `events' -- time periods when a particular regime is strongly expressed. We refer to a collection of events as a `catalogue'. To define the events that make up a catalogue, we first compute the discrimination score \citep{ris21:mwr} that describes the dominance, or otherwise, of a single archetype at a given time step. First, note that due to the convexity constraints on $\mathbf{S}$, the sum over all $p$ at time $t$ is 1 (i.e. $\sum_{k}^{p} S_{kt} = 1$). Intuitively, if the affiliation probability for a single archetype $k$ approaches 1 at time $t$, then the affiliation probability for all other archetypes $\neq k$ must approach zero, and those regimes are very unlikely to be expressed at that moment. Conversely, where all affiliation probabilities distributed equally amongst all archetypes (and hence equal to $1/p$), no archetype can be considered dominant, and it is unlikely that the snapshot at that time resembles any particular archetypal pattern. \newline

%{\color{red}VL: I would say $\approx1$ and $\approx 1/p$, %respectively. The cases indicated above have zero probability of %occurrence!}

Taking advantage of this intuition, we form the discrimination score as in \citet{ris21:mwr}:
%\begin{linenomath*}
\begin{equation} \label{Eqn:Discrimination_score}
\Delta_p(t) = 1-\left (\frac{1}{p-1} \right) \left ( \frac{1}{S_{p_{\textrm{max}}t}}-1 \right ),
\end{equation} 
%\end{linenomath*}
where $p_\textrm{max}$ is the \textit{winning} archetype with the highest affiliation probability at time $t$. The discrimination score lies between 0 -- ie. the affiliation timeseries have equal values of $1/p$ for all archetypes and there is hence no discrimination -- and 1 -- a single archetype with an affiliation equal to 1 while all others have an affiliation value of 0. \newline

We use the discrimination score, together with a persistence criterion to identify events is as follows: 
\begin{itemize}
\item identify periods for which a single archetype is `winning': that is has an affiliation greater than all the others;
\item of those periods, determine periods where the discrimination score is greater than 0.8;
\item determine a block event: onset is defined as the day when the discrimination score passes from less than 0.8 to greater than 0.8, while termination is the day when the discrimination score drops below this threshold. The discrimination score may drop below the threshold for a single day before recovering and still be defined as a single event. Event duration is the defined the number of days between onset and termination;
\item events with a duration of greater than 5 days are added to the catalogue.  
\end{itemize} 
The choice of thresholds for both discrimination score and persistence is, of course, somewhat arbitrary. We have tested many different parameters, finding that the number of events is far more sensitive to persistence than discrimination score. The combination of $\Delta=0.8$ and a persistence of 5 days provides a good trade-off between a too many events to capture persistent extremes, and too few for adequate statistical power. We will discuss these points further in the context of return periods in Sec. \ref{Sec:Discussion_Conclusion}.  

\subsubsection{Rossby Wave Packet Detection}% \label{Sec:RWP_detection}

RWPs are generally defined as a finite number of troughs and ridges arising from Rossby waves with a spatially and temporally varying amplitude \citep{WirthEtAl2018}. This amplitude, or envelope, modulates the magnitude of perturbations to the climatological background state, with larger spatial and temporal scales than the individual peaks and troughs contained therein. Energy is typically passed from individual peaks or troughs to their immediate downstream neighbour, in a process known as downstream development that ultimately manifests as the propagation of the whole packet at the group velocity. \newline

Here, we identify RWPs by complex demodulation via Hilbert transforms \citep{Titchmarsh1948}, as described in \citet{ZiminEtAl2003} with a similar procedure used in \citet{FragkoulidisEtAl2018}:

\begin{itemize}
    \item For each latitude, the 200-hPa meridional velocity anomaly fields are band-pass filtered spatially, removing scales smaller than $\sim$2,000km and larger than $\sim$10,000km. Filtering is performed in the space domain using a finite impulse response Hamming window filter \citep{Oppenheim1999} designed to minimise pass-band ripple. A second order forward-backwards filtering is performed to avoid inducing phase shifts;
    \item The Hilbert transform is applied to the filtered anomalies to obtain the analytic signal;
    \item The absolute value of the analytic signal is taken as the envelope of the 200-hPa meridional velocity anomalies.  
\end{itemize}

\section{Results} \label{Sec:Results}

\subsection{Extreme Regime Patterns for Detrended Data }

AA is applied to detrended JRA-55 reanalysis daily-mean surface temperature anomalies $T_\textrm{2m}$ for the period 1958-2023, over the extend boreal summer (May-September). We compute archetypes from $p=2$ to $p=20$, inclusive, in order to assess the effects of the number of potential regimes identified by the method. Careful assessment of results has lead us decide on using 8 archetypes for the majority of this study - which provides a good trade-off between simplification of the data and discrimination between distinct regimes. A discussion of the sensitivity of the resulting spatial patterns to the number of archetypes is included in Appendix A. \newline

%{\color{red}VL: this needs to be a bit more precise/explicit; maybe showing the archetypes for $p$ say =5, =15 and =20 in the supplementary material}

\begin{figure}[p]
 \centerline{\includegraphics[width=\textwidth, height=7.8in]{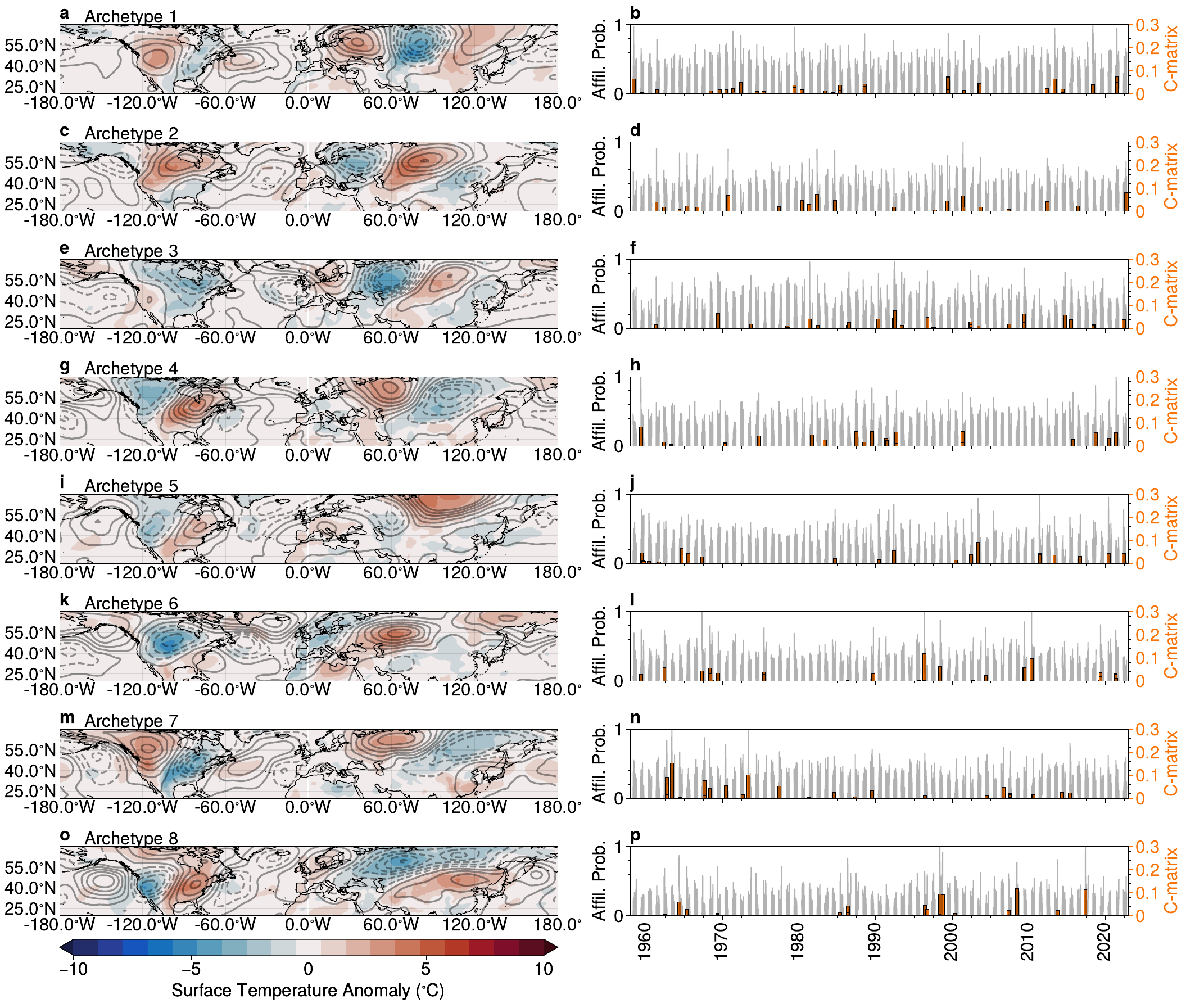}}
  \caption{\textbf{Archetypal Patterns and Affiliation computed from detrend JRA-55 daily maximum surface temperatures anomalies:} \textbf{left panels} spatial patterns for each of the 8 ``regimes" showing $T^{\prime}_\textrm{2m max}$ (colours) and z500 (contour lines). The contour interval is 50m, with a maximum value of 300m, dashed contour lines show \textit{negative} anomalies, while solid contour lines show \textit{positive} anomalies; \textbf{right} affiliation probability (grey lines) and $\mathbf{C}$ matrix weights (orange bars) corresponding to each spatial pattern.}
  \label{fig1:Archetype_Patterns_with_S_and_C_detrend}
\end{figure}

The archetypal patterns for $T_\textrm{2m}$ and 500hPa geopotential are formed by compositing on the $\mathbf{C}$ weights following Eqn. \ref{Eqn:C_composite} are shown in Fig. \ref{fig1:Archetype_Patterns_with_S_and_C_detrend} (left panels), together with the associated affiliation timeseries and $C$-matrix weights. Coherent spatial patterns showing warm surface temperature anomalies are found adjacent to, yet displaced from, high pressure systems for each archetype. Temperature anomalies in Fig. \ref{fig1:Archetype_Patterns_with_S_and_C_detrend} typically exceed 5$^{\circ}$C in the warmest regions, and occasionally approach 10$^{\circ}$C. The archetypes also show a circum-hemispheric structure with two or more distinct warm regions over a wide area. These patterns project mostly on longitudinal wavenumbers 4-7, (see discussion in Section \ref{Sec:Dynamics}). For example, the patterns for Archetype 1 (Fig. \ref{fig1:Archetype_Patterns_with_S_and_C_detrend}\textbf{a}) show warm centers over western North America, western and central Europe, and east Asia. Warm centers over land have larger amplitudes than those over the ocean. \newline

The $\mathbf{S}$ (grey) and $\mathbf{C}$ (orange bars) time series for each archetype are shown in Fig. \ref{fig1:Archetype_Patterns_with_S_and_C_detrend}(right panels). The distribution of the $\mathbf{C}$-matrix weights is instructive, as they show the time snapshots and weights applied that are used to construct the archetypal patterns via Eqn. \ref{Eqn:AA_2}. Take, for example, Archetype 1. The distribution of C-matrix weights has several distinct clusters in time, notably in the late 1960s and early 1970s, the early 2000s, and from 2012-2015. In contrast, the $\mathbf{C}$-matrix weights are zero for the majority of the 1990s, and the mid-2000s. All other archetypes show similar periods of with non-zero $\mathbf{C}$-matrix weights and periods of absence, suggesting that the time snapshots drawn to construct the archetypes are distributed through the data time period. The affiliation probability is correlated with the $\mathbf{C}$-matrix weights, being larger when the $\mathbf{C}$-matrix weights are nonzero. However, it is important to note that the affiliation probability can be high (occasionally approaching 1) even when the $\mathbf{C}$-matrix weights are zero. 
Indeed, this fact gives AA its power, allowing us to assign a affiliation at every time step in the dataset. \newline

%{\color{red}VL: In order to help the reader, can we say something like "C is associated with the learning phase of the algorithm, whilst S uphelds the reduced-order description of the state of the atmosphere".}

\subsection{Links Between Heatwave Events and Extreme Regimes } \label{Sec:Extreme_Events}

\begin{figure}[hp!]
 \centerline{\includegraphics[width=7in]{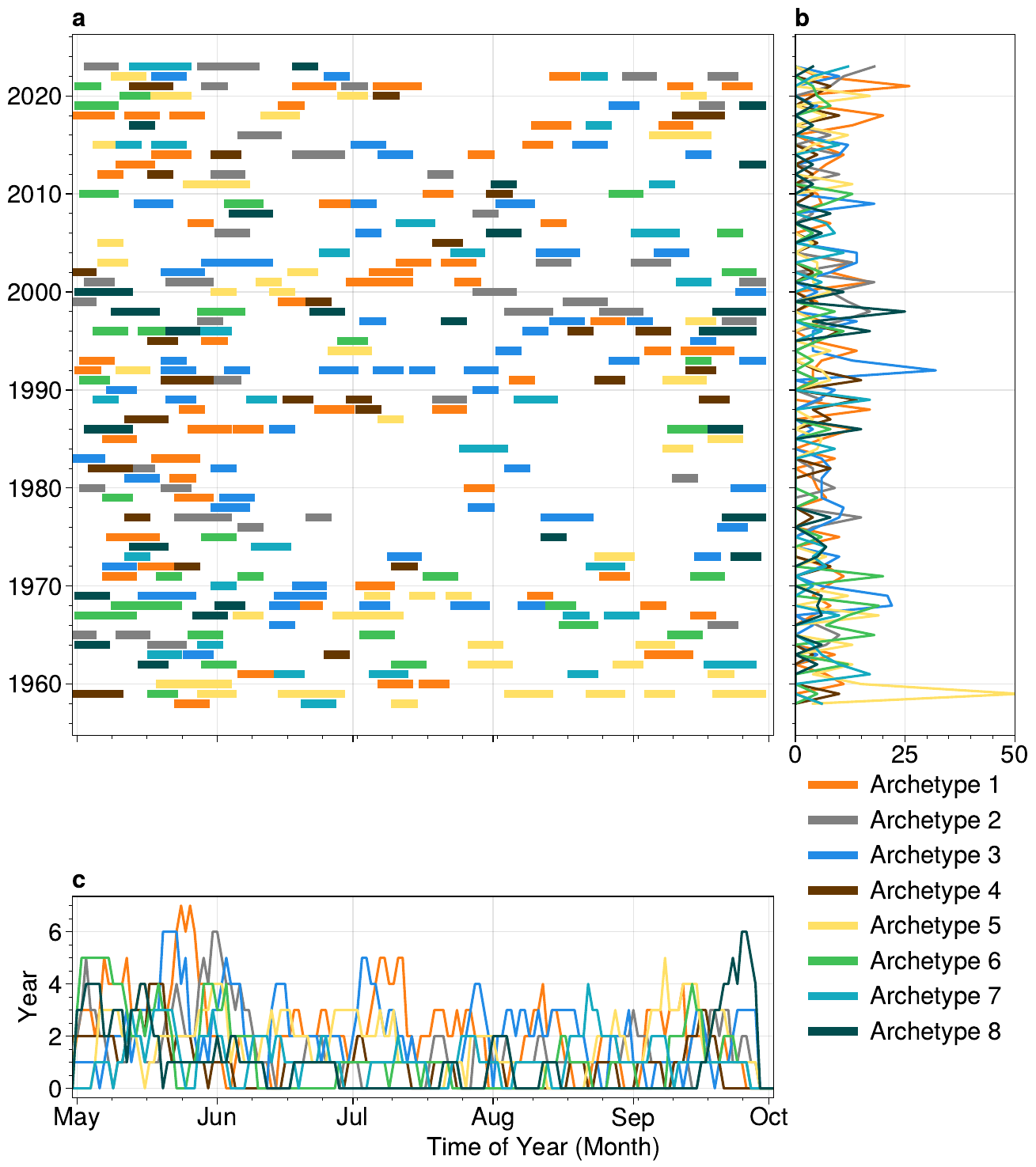}}
  \caption{\textbf{Temporal distribution of events from the automated detection procedure:} \textbf{a} A calendar showing the distribution of events, coloured by the relevant archetype, by year ($y$-axis) and day of year ($x$-axis). White regions show times when no event was detected.; \textbf{b} The total event days for each regime occurring in each year; \textbf{c} The annual cycle for the total event days per year, for each regime. The colour legend is shown in the bottom right of the page. The method for specifying events is described in section \ref{Sec:Methods_and_Data}\ref{Sec:Methods}\ref{Sec:Event_Identification}}
  \label{fig2:Event_Lego_Plot_detrend}
\end{figure}

Using the eight archetypes discussed above, we form an event catalogue (described in Section \ref{Sec:Methods_and_Data}\ref{Sec:Methods}-\ref{Sec:Event_Identification}), shown diagrammatically in Fig. \ref{fig2:Event_Lego_Plot_detrend}. Events of particular archetype tend to cluster, showing intra-decadal and decadal variability. For example, events associated with Archetype 1 (orange) were particularly common between July and August between the years 2001 and 2003, while events associated with Archetype 5 (yellow) were common during the period 1958-1970, were almost absent for the following 20 years, and then somewhat more common after the year 2000. As another example, a period between 1970 and 1990 shows a drammatically reduced number of events between June and September, when compared to other 20 year periods. By simply counting the number of event days for each archetype in every year (Fig. \ref{fig2:Event_Lego_Plot_detrend}\textbf{b}),  it is apparent that occasionally, a single archetype will be particularly common. For example, Archetype 1 can be seen to reach 25 event days in 2021, while Archetype 5 reached 50 event days in 1958. In contrast, the evidence for a regular seasonal cycle is more limited, as shown in Fig. \ref{fig2:Event_Lego_Plot_detrend}\textbf{c}. With the exception of a peak in the number of event days between May and June that fails to reach statistical significance, no clear seasonal cycle can be determined from the data, which suggests that these regimes can occur with roughly equal probability any time during the extended boreal summer. \newline

\begin{figure}[p!]
 \centerline{\includegraphics[width=7in, height=7.in]{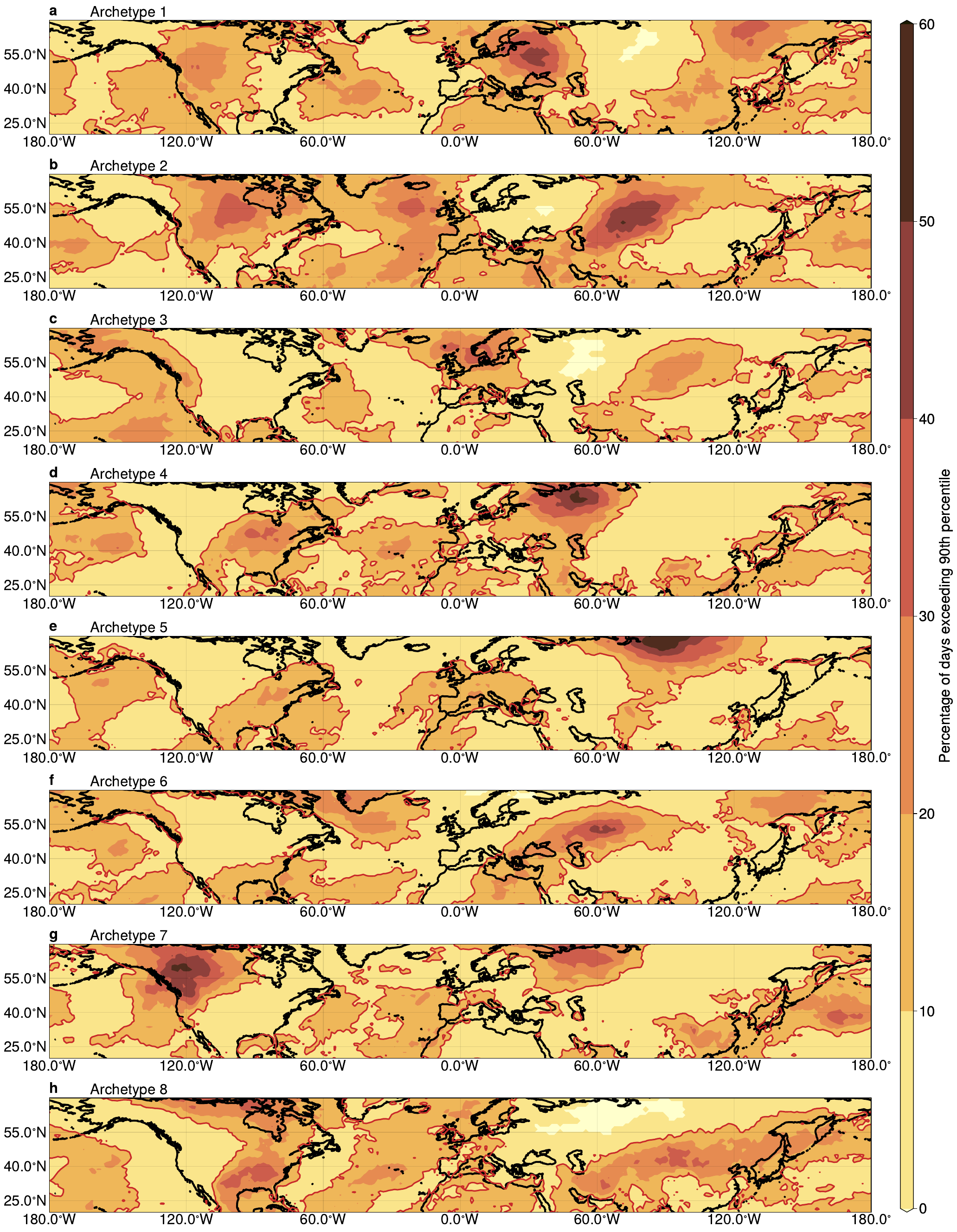}}
  \caption{\textbf{The percentage of event days that exceed the surface temperature 90th percentile}: For each of the events associated with each individual archetypal pattern in the catalogue, the number of days that exceed the 90th percentile, divided by the total number of event days. The events associated with archetypal patterns 1 through 8 are shown down the page. Solid red contour shows the 10\% level, which we might expect if days were drawn randomly.}\label{fig4:Exceed_90th_Percentile}
\end{figure}

In Fig \ref{fig4:Exceed_90th_Percentile} we demonstrate that these events are typically associated with large and coherent areas of extreme temperatures by determining the percentage of event days that exceed the 90th percentile of $T_{2m}$ for the month in which each event day falls. Distinct regions where a high percentage of event days have extreme temperatures are evident for all archetypal patterns. Each archetype generally shows one coherent region where the percentage of event days exceeding the 90th percentile is greater than 50\%, as well as one or two secondary regions where the percentages exceed 30\% or 40\%. As an example, for events associated with Archetype  1 (Fig. \ref{fig4:Exceed_90th_Percentile}\textbf{a}), more than 50\% of event days exceed the 90th percentile over a region centered on central Europe between longitudes of 0$^{\circ}$ and 60$^{\circ}$E. Additionally, archetype 1 type events also include regions in western North America (between longitudes 120$^{\circ}$W and 90$^{\circ}$W) and northeast Asia (between longitudes 80$^{\circ}$E and 140$^{\circ}$E) where 30 to 40\% of event days exceed the 90th percentile. When we compare the spatial patterns in Fig. \ref{fig4:Exceed_90th_Percentile} with the corresponding patterns of $T_{2m}$ in Fig. \ref{fig1:Archetype_Patterns_with_S_and_C_detrend} we see, as might be expected, a correspondence between regions of positive temperature anomalies in the archetypcal patterns and those regions likely to experience extreme temperatures during events associated with that archetype. \newline

\subsubsection{Individual Case Studies}

\begin{figure}[p!]
 \centerline{\includegraphics[width=7in, height=5in]{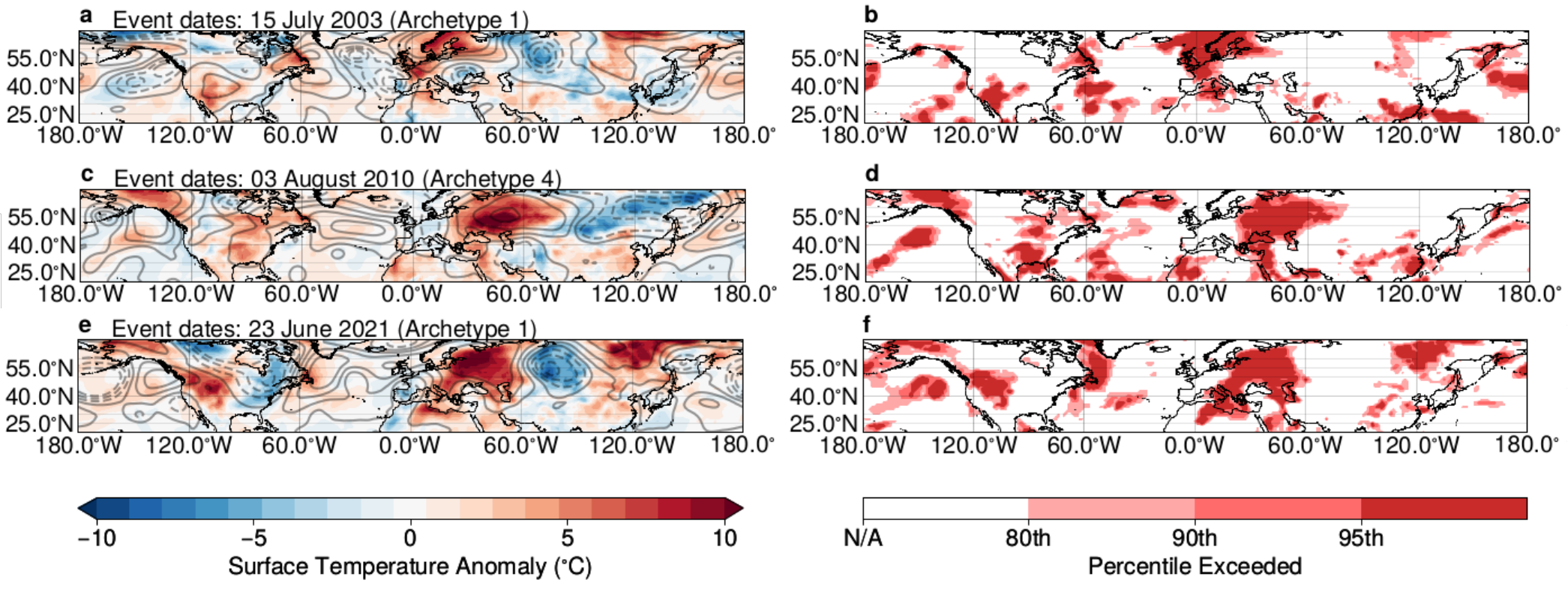}}
  \caption{\textbf{Three example events from the automated detection procedure:} \textbf{left} spatial patterns of $T_\textrm{2m max}$ (colours) and z500 (contour lines) for the date falling halfway between event onset and termination; and \textbf{right} areas where the surface temperature anomaly exceeds the 80th, 90th or 95th percentile. The date for each plot is indicated in the in-panel text. These dates correspond to (\textbf{a,b}) the 2003 western European (French) heatwave; (\textbf{c,d}) the 2010 central European (Russian) heatwave; and the 2021 western North American "heatdome" heatwave.  }\label{fig3:Canonical_Events_from_Catalogue}
\end{figure}

We now investigate the physical manifestation of individual events in the AA catalogue through a set of individual case studies. The events chosen are three heatwaves discussed extensively in both the scientific literature and popular press: the July-August 2003 western European heatwave \citep{BlackEtAl2004}, the July-August 2010 central European or ``Russian" heatwave \citep{FragkoulidisEtAl2018}, and the June-July 2021 western North American ``heatdome" \citep{BartusekEtAl2022,WhiteEtAl2023,LucariniEtAl2023}. For each event, we plot the detrended surface temperatures and 500 hPa geopotential anomalies on the `central day', midway between the start and end dates in the left hand column of Fig. \ref{fig3:Canonical_Events_from_Catalogue}. In all cases studies presented here, large areas of anomalously high temperatures, locally exceeding 10$^{\circ}$C, were accompanied by large blocking high pressure systems. We also note that the 2003 and 2021 events are both periods dominated by the same regime -- that associated with Archetype 1.  \newline

To place these anomalies in the historical context, the right hand panels of Fig. \ref{fig3:Canonical_Events_from_Catalogue} show regions where the temperature anomalies exceeds the 80th, 90th and 95th percentile, which clearly indicate coherent high temperature anomalies at the synoptic to continental-scale. For example, the event that occurred in August 2003 (Fig. \ref{fig3:Canonical_Events_from_Catalogue}\textbf{a,b}). On the date shown, temperatures exceed the 95th percentile through much of the Mediterranean and western Europe, as well as the southwestern North America. The August 2010 event, shown in Fig. \ref{fig3:Canonical_Events_from_Catalogue}\textbf{c,d},  temperatures greater than the 90th and 95th percentile span 35-40$^{\circ}$ of longitude over much of Central and Eastern Europe, Central Asia, and a separate region of high temperatures over eastern North America. The June 2021 event, shown in Fig. \ref{fig3:Canonical_Events_from_Catalogue}\textbf{e,f}, shows extensive regions where temperatures exceed the 90th or 95th percentile over western of North America, and central and eastern Europe, as well as northeast Asia. Indeed, the region of extremely high temperatures over Europe  has a greater spatial extent than that over North America. \newline

From the analysis presented above, we make two inferences. The first is that events selected from our semi-automated catalogue correspond to situations with large-scale coherent heatwave conditions, with surface air temperatures exceeding the 90th or even 95th percentile over vast regions. Secondly, the large-scale patterns selected by AA produce compound and concurrent events, with heatwave conditions occurring simultaneously in two or more distinct parts of the globe. \newline

\subsection{Representation of an event by a single extreme regime} \label{Sec:Single_Archetype}

\begin{figure}[p]
 \centerline{\includegraphics[width=7in, height=5in]{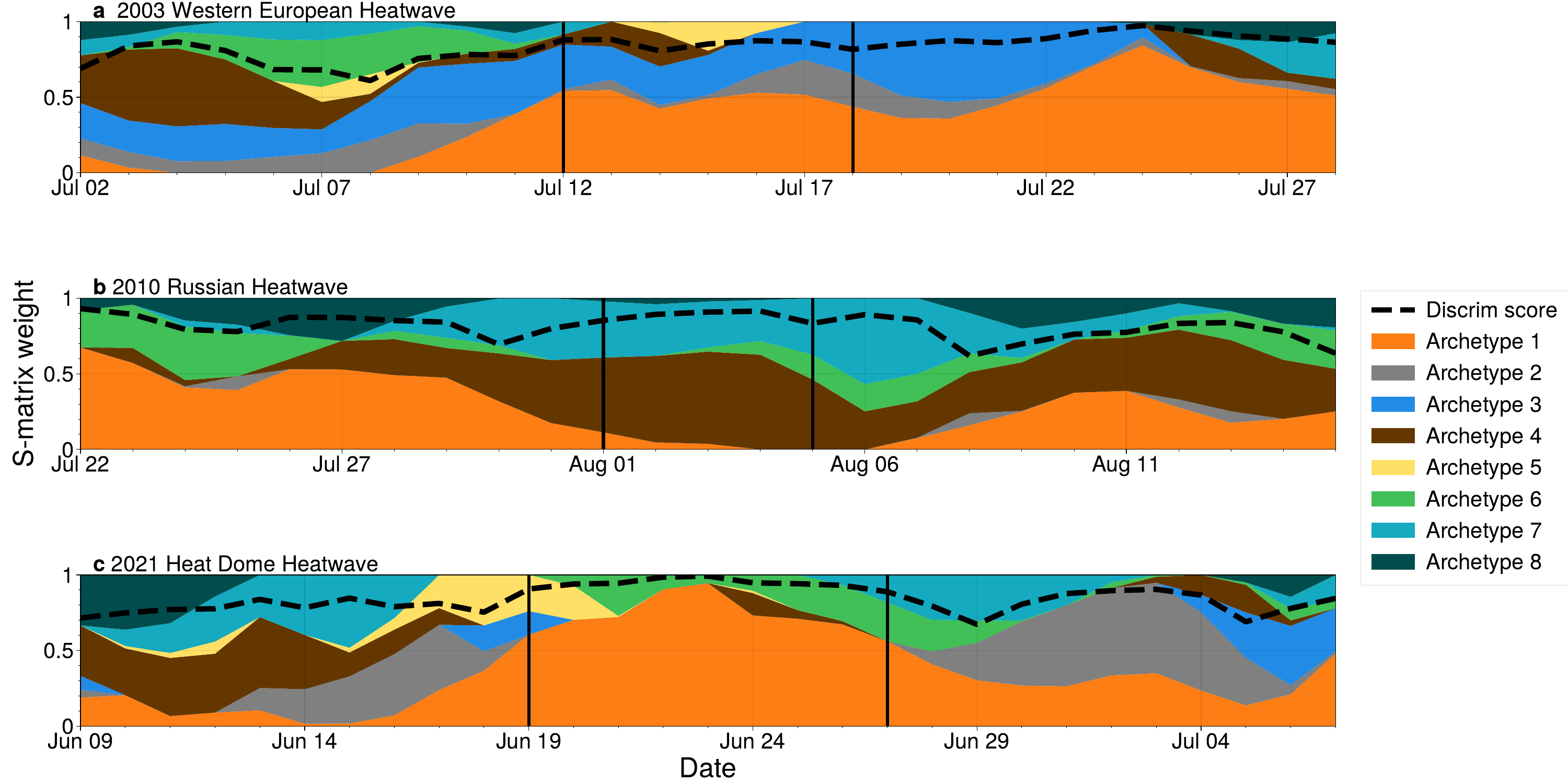}}
  \caption{\textbf{Affiliation probabilities during case studies} Stacked affiliation probabilities (ie. the $\mathbf{S}$-matrix weights) during the \textbf{a} 2003 Western European (French) heatwave, \textbf{b} the 2010 ``Russian" heatwave; and \textbf{c} 2021 ``Heatdome" heatwave. The time period show commences 10 days prior to event onset, and continues 10 days post event termination, as indicated by the solid back vertical lines.        }\label{fig6:Stacked_Smatrix_canonical_events}
\end{figure}

Our catalogue explicitly selects events where a single `winning' archetype is strongly expressed, as described in Sec. \ref{Sec:Methods_and_Data}\ref{Sec:Methods}--\ref{Sec:Event_Identification}.  We have shown that these regimes tend to favour extreme surface temperatures. Our event identification algorithm does not select explicitly for large deviations from climatology, but instead for the strong and persistent expression of a single archetypal pattern. However, AA allows for the expression of the data as a mixture of two or more different archetypes. To demonstrate, in Fig. \ref{fig6:Stacked_Smatrix_canonical_events} we show a stacked plot of the affiliation probability (the $\mathbf{S}$-matrix weights), along with the discrimination score $\Delta_{8}$ (Eqn. \ref{Eqn:Discrimination_score}) for each of the three case studies described above, as well 10 days both pre-and-post event. In the lead up to each event, multiple archetypes are expressed with no single state clearly dominant. During the events, a single archetype dominants with affiliation weights that exceed 0.5. and $\Delta_{8}>0.8$ The 2003 and 2021 events  (Fig. \ref{fig6:Stacked_Smatrix_canonical_events}\textbf{a,c}) were dominated by Archetype 1, while the 2010 Russian heatwave  (Fig. \ref{fig6:Stacked_Smatrix_canonical_events}\textbf{b}), was dominated by Archetype 4. However, during the 2003 event Archetype 3 is also expressed, while Archetype 7 is expressed during the 2010 event. As such, even with a procedure that explicitly selects for the strong expression of a single archetype, the resulting representation of the surface temperature from AA is likely to include the influence of multiple regimes. \newline

 To quantify the effects of defining an event by a single archetype, we note that Eqn. \ref{Eqn:AA_min}, while being used to define the AA problem, can also be used to reconstruct the original data from either all or a subset of the archetypal patterns once the $\mathbf{C}$ and $\mathbf{S}$ weights are known, through the expression: 
    \begin{equation} \label{Eqn:AA_reconstruction}
        \Tilde{\mathbf{X}}_{s \times t} = \mathbf {X}_{s \times t}\mathbf{C}_{t \times \tilde {p}} \mathbf{S}_{\tilde {p} \times t},
    \end{equation}
where $\tilde{p}=1,\dots,P$ is a index of a single archetypal pattern. Essentially, Eqn. \ref{Eqn:AA_reconstruction} acts as a kind of filter, allowing for any time step in the original dataset to be reconstructed by temporally weighting the spatial patterns shown in Fig. \ref{fig1:Archetype_Patterns_with_S_and_C_detrend}. \newline

\begin{figure}[p]
\centerline{\includegraphics[height=5.0in,width=7.0in]{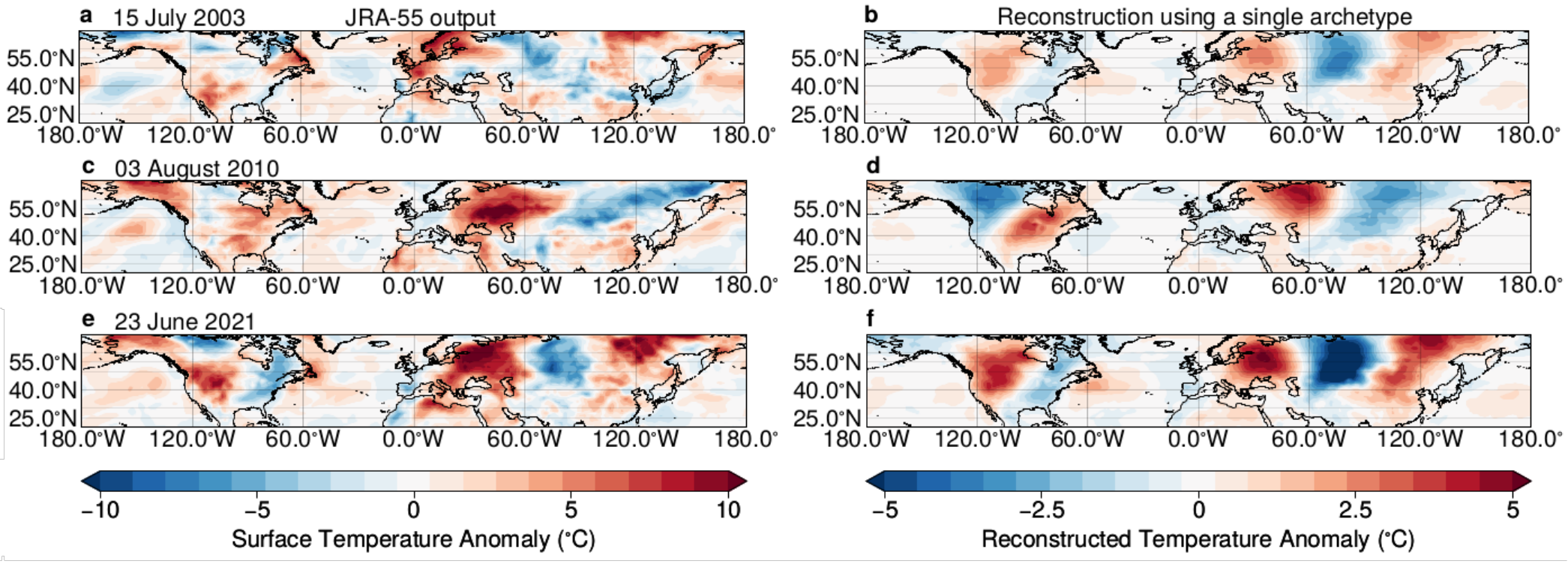}}
  \caption{\textbf{Reconstruction of time snapshots from a single archetype} (left column) $T_{\textrm{2m}}$ anomalies and (right column) reconstructed $T_{\textrm{2m}}$ anomalies from the single `winning' archetypes, for the three event case studies: \textbf{a,c} the July 2003 Western European (French) heatwave; \textbf{c,d} the August 2010 Russian heatwaves; and \textbf{e,f} the June 2021 North American `heatdome'. Note that the range of the reconstructed fields is a factor of 2 smaller than the full fields. } \label{fig7:Canonical_Events_from_reconstruction}
\end{figure}

In Fig. \ref{fig7:Canonical_Events_from_reconstruction}, we show both the $T_{\textrm{2m}}$ anomalies alongside those reconstructed from the dominant archetype for single day snapshots during each of the case studies. The reconstructed fields show similar continental-scale features to those from the complete dataset. For the 2003 and 2021 case studies, reconstructed from Archetype 1, show three broad and coherent regions of warm anomalies over western North America, western Europe, and east Asia, while the 2010 event shows two warm regions, over central Europe and eastern North America. The reconstructed patterns are in rough agreement with the snapshots taken from the complete dataset, although we note that the latter contain, as expected, smaller-scale structures not present in the reconstructions, and that there are significant differences in the exact placement of the warm anomalies. For example, the 2003 event shows warm temperatures further to the west over Europe and further to the south over North America in the full dataset than in the reconstruction. We also note that the magnitudes of the reconstructed fields are typically about 50\% of the full fields, which occurs since AA, unlike more common matrix decomposition methods such as PCA, does not preserve variance. Including more archetypes increases the magnitude of the reconstructed anomalies and tends to the reconstructed and full fields into greater agreement, although the effect is minor as each event has been chosen for the dominance of a single archetype. \newline

\begin{figure}[p]
\centerline{\includegraphics[height=6.0in,width=6.0in]{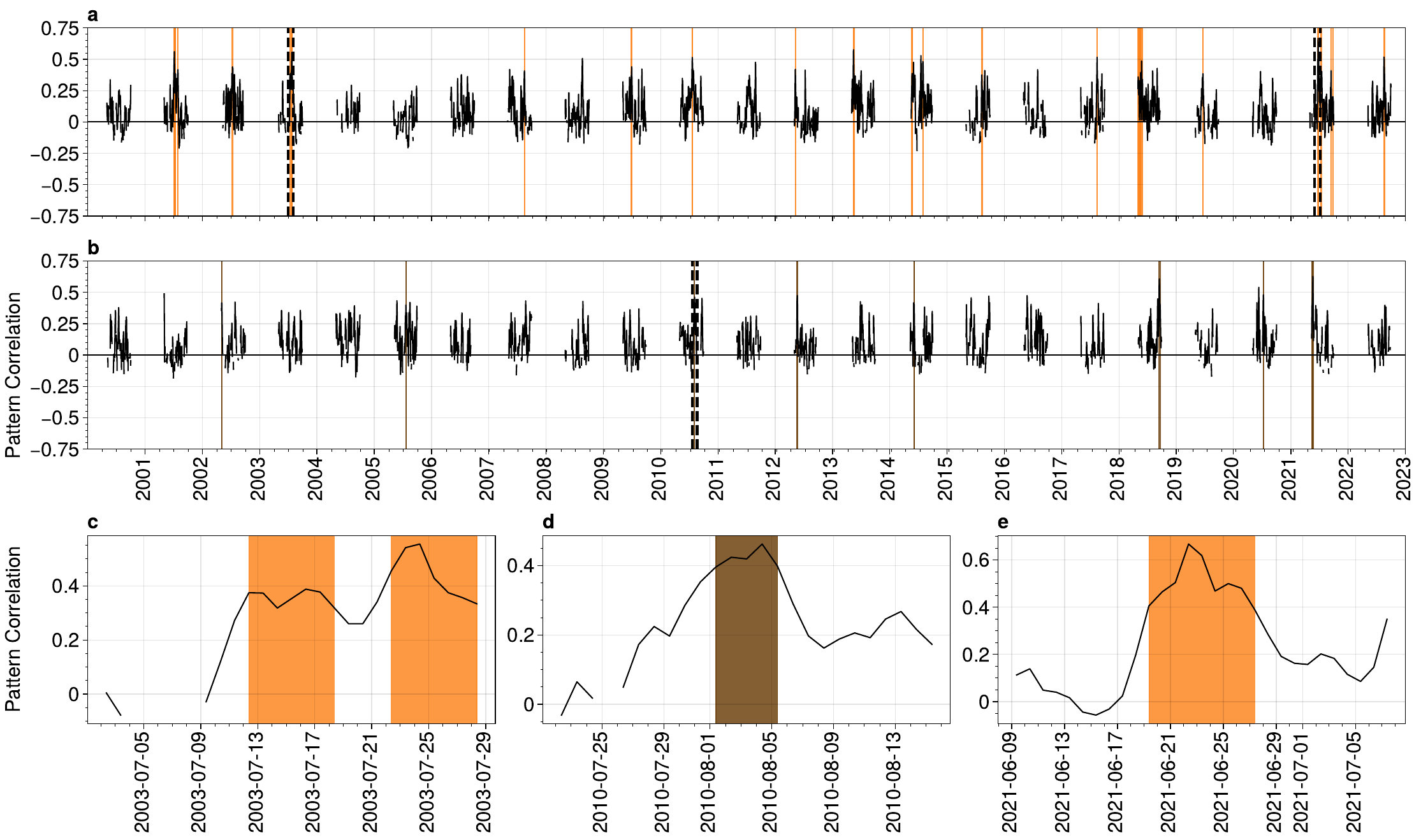}}
  \caption{\textbf{The similarity of the atmospheric state to the archetypal patterns through time}: The pattern correlation for the $T_{\textrm{2m}}$ at any time snapshot and \textbf{a}) archetypal pattern 1; or \textbf{b}) archetypal pattern 4, computed over the entire northern hemisphere. Events from the archetype derived catalogue are shown as thin shaded vertical lines. Panels \textbf{c}), \textbf{d}), and \textbf{e}) are zoomed to show periods corresponding to the July 2003 Western European (French) heatwave; the August 2010 Russian heatwave; and the June 2021 `heatdome' event.   } \label{fig8:Pattern_Correlation_vs_events}
\end{figure}

To quantify the differences between the reconstructed fields and the complete dataset, we compute the Pearson pattern correlation between the JRA-55 $T_{\textrm{2m}}$ 
anomalies and those reconstructed from a single archetype. The pattern correlation is defined as the Pearson $r_{t}$ over space, as in \cite{RisbeyEtAl2023}:
    \begin{displaymath} 
        r_{t}=\frac{ \sum_{s} (\Tilde{\mathbf{X}}_{s \times t} - \overline{\Tilde{\mathbf{X}}_{t}}) (\mathbf{X}_{s \times t} - \overline{\mathbf{X}_{t}}) } { \sqrt{ \sum_{s} (\Tilde{\mathbf{X}}_{s \times t} - \overline{\Tilde{\mathbf{X}}_{t}} )^{2} } \sqrt{ \sum_{s} (\mathbf{X}_{s \times t} - \overline{\mathbf{X}}_{t} )^{2}} }
    \end{displaymath}
where $\overline{\mathbf{X}_{t}}$ and $\overline{\Tilde{\mathbf{X}}_{t}}$ are the averages over all space points at every time step. Fig. \ref{fig8:Pattern_Correlation_vs_events} shows the pattern correlation between the reconstructed and full fields over the entire Northern Hemisphere for every time step in the database, for Archetypes 1 and 4. Events, shown as shaded regions, tend to align with pattern correlation maxima, although we note that the value of $r_{t}$ rarely exceeds 0.5. To show this more clearly, we zoom to a one month period around the 2003, 2010 and 2021 case studies. In each event, we see a clear peak in the pattern correlation during the event periods, with values that reach a maximum of  $\sim$0.55 during the 2003 event, $\sim$0.45 during the 2010 event. and $\sim$0.7 during the 2021 event. \newline

These results demonstrate the advantages and disadvantages of defining events by their large scale structure. It is clear that the archetypal patterns show a degree of correlation with the complete data field during events taken from our catalogue, typically exceeding 0.4 and occasionally exceeding 0.7, across the entire northern hemisphere. The fact that temporal peaks in $r_{t}$ occur during each of our events suggests that there is some merit in describing regional extremes as the result of the strong expression of a few large-scale configurations coming from the boundaries of the attractor. However, while the continental-scale patterns are undoubtedly important, smaller-scale features are of great importance at the regional-scale, and are not captured by a single archetype. The analysis of the pattern correlation indicates that event during the events, upwards of 50\% of the variance across the northern hemisphere remains unexplained by a single regime. Including all archetypal patterns in the reconstruction, as opposed to just a single archetype (not shown), can increase the pattern correlation to above 0.8 for certain periods, although the increase in $r_{t}$ is much smaller during events as these are, by definition, periods when a single archetype is dominant. \newline

\section{Dynamical Interpretation of AA Identified Regimes} \label{Sec:Dynamics}

We have shown that AA applied to detrended surface temperature anomalies extracts circum-hemispheric patterns that, when strongly expressed, lead to strongly anomalous surface temperatures over broad, distinct regions that may be separated by many thousands of kilometers. High temperatures are almost always associated blocking high pressure systems.  Previous work has shown that regional extremes are frequently embedded within quasi-stationary, circum-hemispheric meandering atmospheric circulation with a peak in the wavenumber spectrum between 4 and 8 \citep{KornhuberEtAl2017,KornhuberEtAl2019,YangEtAl2024}. We now build upon these studies, as well as the results presented in previous sections, to propose a plausible dynamical origin for these large-scale patterns, which is one of the main advantages of our outside-in approach. \newline

\begin{figure}[p!]
\centerline{\includegraphics[width=6.5in, height=8.0in]{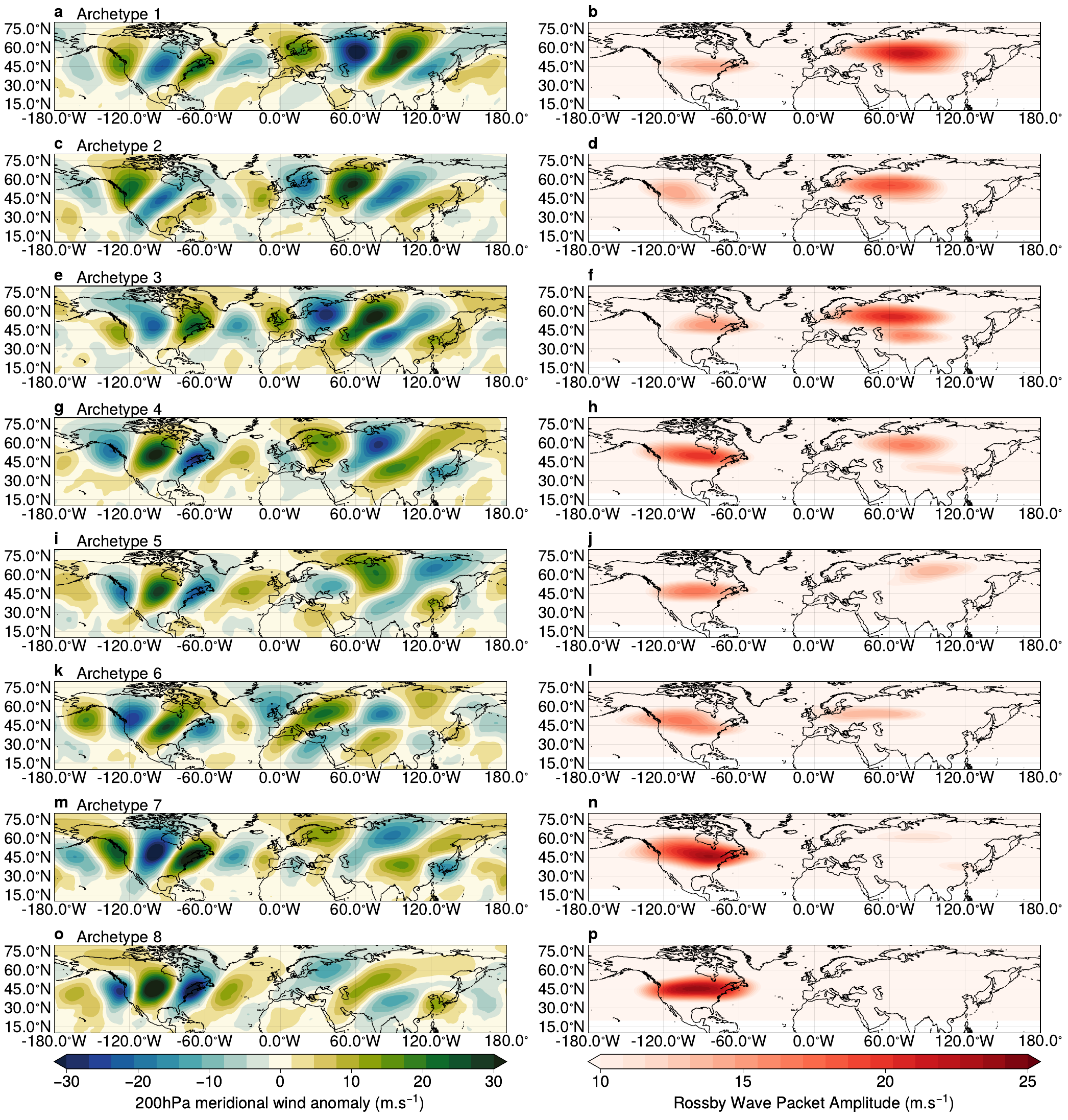}}
  \caption{\textbf{Large scale waves associated with atmospheric circulation associated with archetypal patterns}, $\mathbf{S}$-matrix composites (Eqn. \ref{Eqn:S_composite}) of (left column, \textbf{a,c,e,g,i,k,m,o}) 200hPa de-trended meridional wind velocities; and (right column, \textbf{b,d,f,h,j,l,n,p}) Rossby Wave Packet (RWP) amplitude.    }   \label{fig8:v200_S_composites}
\end{figure}

To begin, we compute the affiliation-weighted composite averages (Eqn. \ref{Eqn:S_composite}) of the 200 hPa meridional velocity anomaly, representing the mid-latitude upper-tropospheric circulation, which is shown in Fig. \ref{fig8:v200_S_composites} (left panels). Each archetype shows a wavy, meandering circulation, characterised by alternating positive and negative anomaly centers. In all archetypes, the wavy patterns span the entire hemisphere. However, there are distinct differences between the archetypes in the location of the anomaly centers, and their number. To quantify the difference in the number of anomaly centers between archetypes, we compute the wavenumber spectrum using a Welch periodogram, computed along latitude 45$^{\circ}$N which approximately follows the anomaly centers. The energy of the spectrum for each Archetypal pattern is generally concentrated into a single, albeit broad peak at either wavenumber 5 (archetypes 1, 3, 4, 5, and 8) or wavenumber 6 (archetypes 2,6,7). However, it is important to note that the energy is broad band, and generally distributed between wavenumbers 4 to 7. \newline  

It is notable that circum-hemispheric meandering patterns shown in Fig. \ref{fig8:v200_S_composites} are modulated, with large amplitudes localised to particular longitudes. For example, the 200hPa meridional wind anomalies associated with Archetype 1 show the strongest amplitudes over North America and central Asia, and the weakest anomalies over the ocean basins. The longitudinal modulation of these signals in the upper-troposphere is characteristic of RWPs \citep{WirthEtAl2018}. To demonstrate this, we compute the affiliation weighted composite of the RWP amplitude (see Section \ref{Sec:Methods_and_Data}\ref{Sec:Methods}-\ref{Sec:RWP_detection}), shown in the right-hand panels of Fig. \ref{fig8:v200_S_composites}. It is clear that the circum-hemispheric wave patterns tend to cluster into two major wave packets located over each of the major land-masses. However, the longitude and amplitude of these packets differs depending on the archetype. As we move from Archetype 1 to Archetype 8, there is a general reduction in the magnitude of the wave packet over Eurasia, and a general increase in the magnitude of the packet over North America. For example, Archetypes 7 and 8 show either very small RWP amplitudes over Europe, which is reflected by relatively low magnitudes of the meridional wind anomalies. When we compare the surface temperature fields associated with each archetype (i.e. Figs. \ref{fig1:Archetype_Patterns_with_S_and_C_detrend} and \ref{fig4:Exceed_90th_Percentile}), we see that the highest anomalies are typically located upstream (that is, westward) of the RWPs, with the exception of Archetype 8, which shows zonally elongated warm anomalies over southern and east Asia that don't appear to be strongly related to the modulation of the wave patterns. As our archetypes are ordered from most to least commonly expressed, we infer that relatively strong RWPs over central Europe and Asia, concurrent with weaker RWPs over North America, are the most common large-scale regime identified here.  \newline

\begin{figure}[p!]
\centerline{\includegraphics[width=6.5in, height=7.5in]{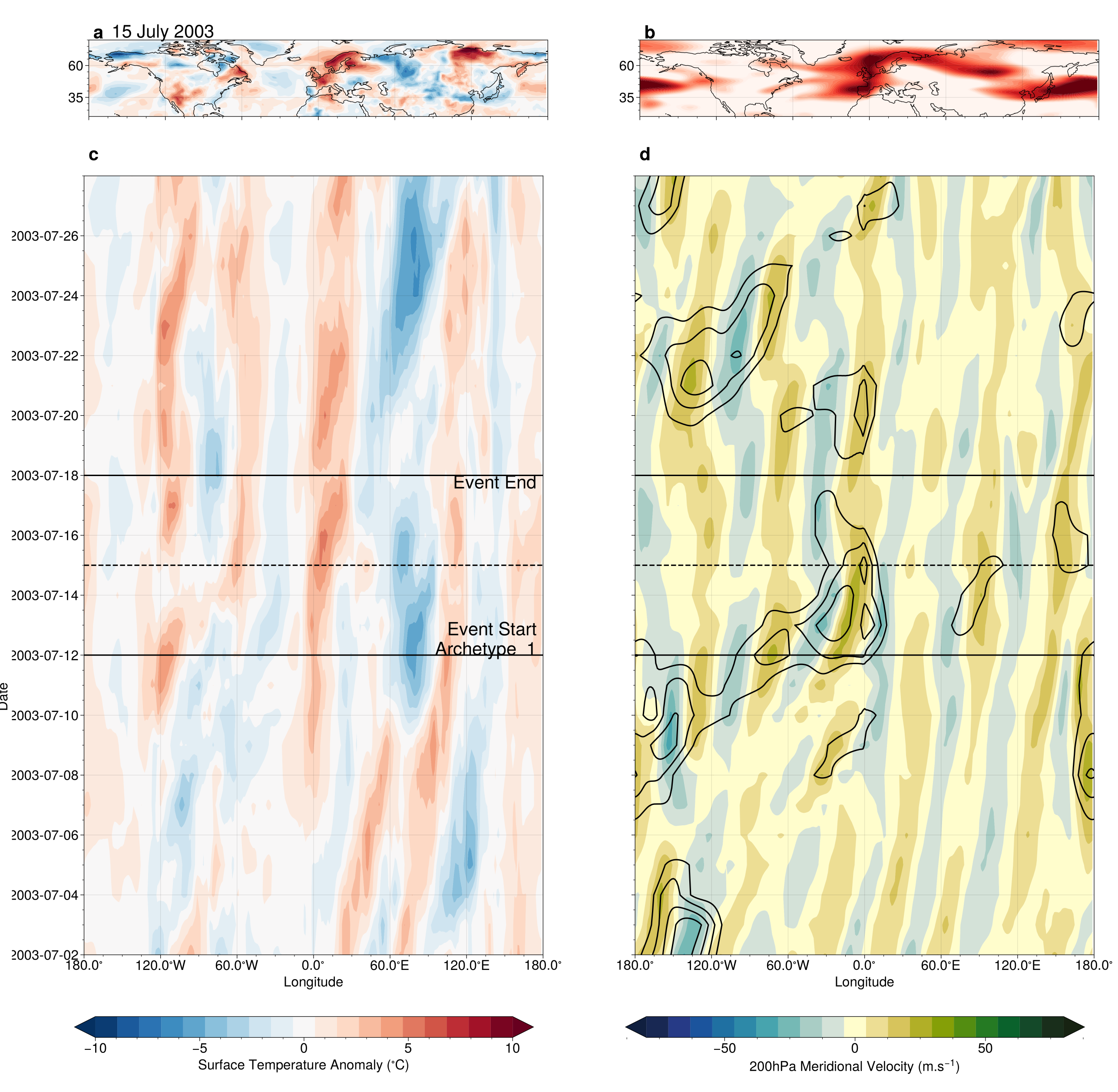}}
  \caption{\textbf{Temporal Evolution of the 2003 ``Western European" or ``French" heatwave}, \textbf{a}) The surface air temperature anomalies; and \textbf{b}) RWP amplitudes on the indicated date. \textbf{c}) Longitude/time (Hovm\"{o}ller) diagrams of  \textbf{c}) $T^{\prime}_\textrm{2m max}$; and \textbf{d}) 200hPa meridional velocity anomalies $v$ (contour lines) and RWP amplitude (C.I. 4 m.s$^{-1}$) commencing 10 days prior to the event onset and until 10 days post event termination. All events and their archetypes are indicated by the horizontal black lines and in figure text. The date for the maps shown in panels \textbf{a}) and \textbf{b}) are indicated by dashed lines.} 
  \label{Fig5:Canonical_Events_case_study_2003}
\end{figure}

\begin{figure}[tp!]
\centerline{\includegraphics[width=6.5in, height=8.0in]{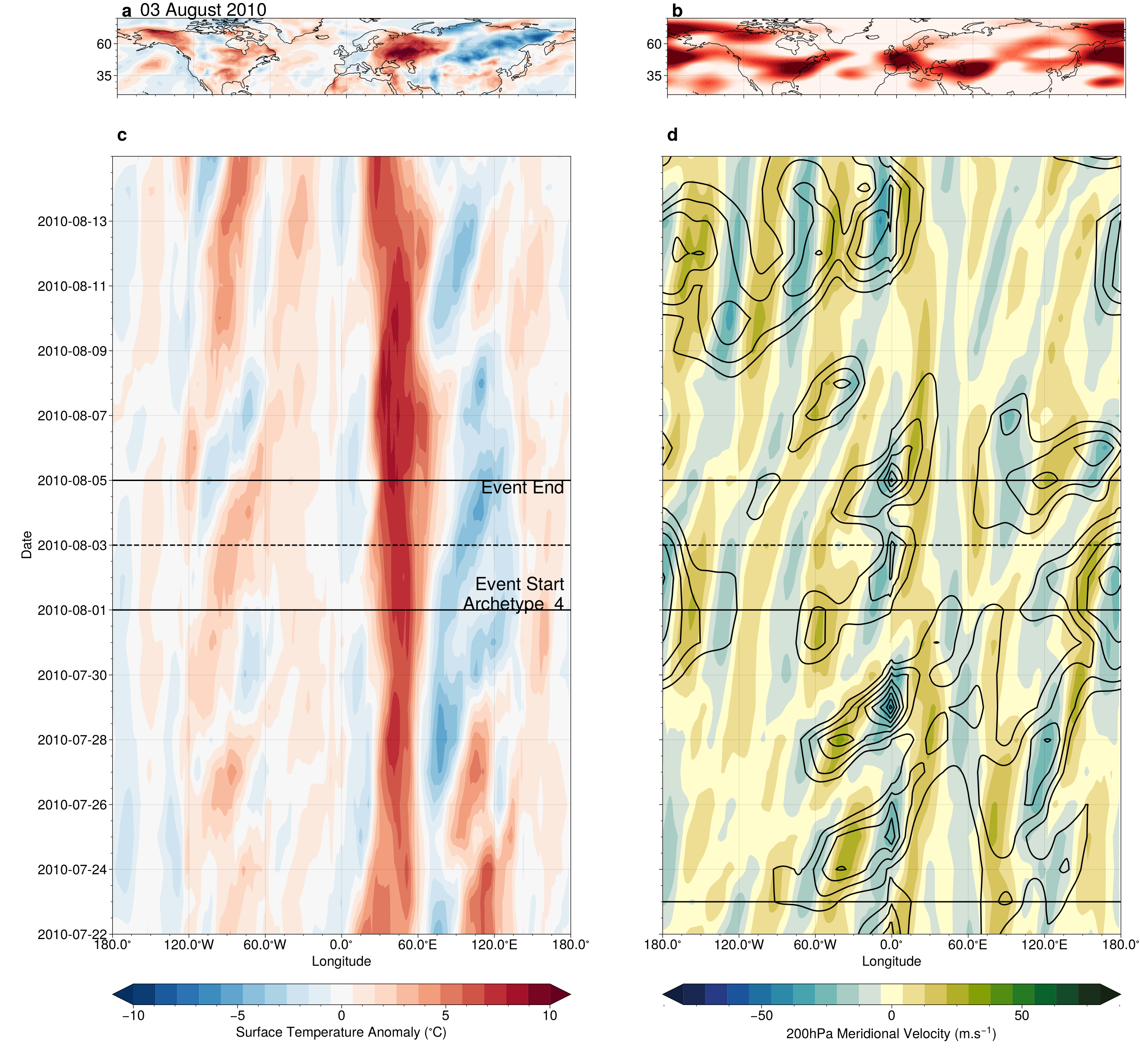}}
  \caption{\textbf{Temporal Evolution of the 2010 ``Russian" heatwave} As in Fig. \ref{Fig5:Canonical_Events_case_study_2003} but for the August 2010 event. } \label{Fig6:Canonical_Events_case_study_2010}
\end{figure}

\begin{figure}[p!]
\centerline{\includegraphics[width=6.5in, height=7.0in]{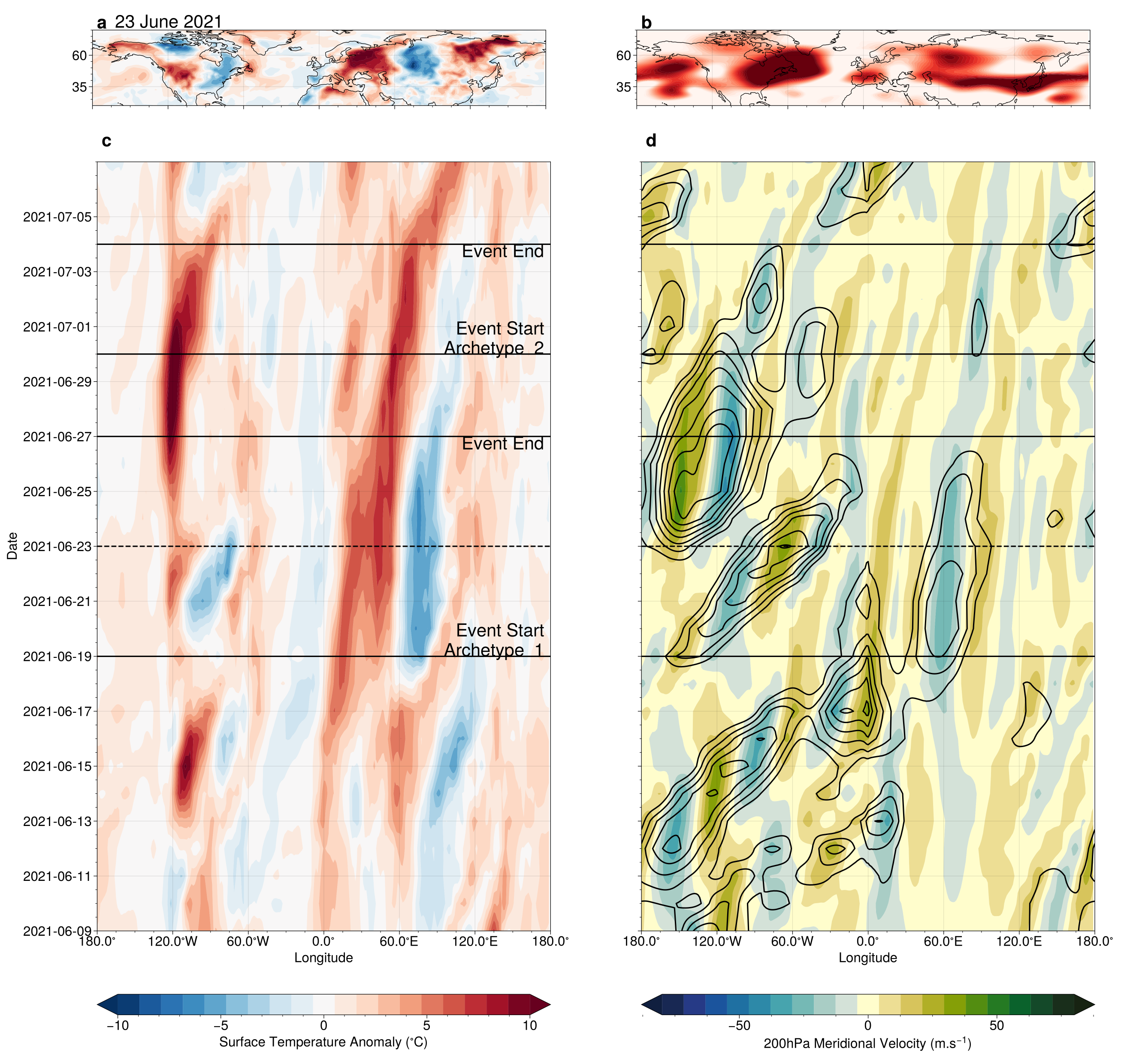}}
  \caption{\textbf{Temporal Evolution of the 2021 Western North-America ``heatdome" heatwave} As in Fig. \ref{Fig5:Canonical_Events_case_study_2003} but for the June 2021 event.}
  \label{Fig7:Canonical_Events_case_study_2021}
\end{figure}

We now investigate how these circum-hemispheric patterns manifest during the case studies. Hence, in Figs. \ref{Fig5:Canonical_Events_case_study_2003}, \ref{Fig6:Canonical_Events_case_study_2010}, and \ref{Fig7:Canonical_Events_case_study_2021} we follow \citet{FragkoulidisEtAl2018} and plot longitude/time (Hovm\"oller) diagrams showing the spatio-temporal evolution of the anomalous surface temperature, 200hPa velocity, and the RWP amplitude averaged between 35$^{\circ}$N and 55$^{\circ}$N, for the three case studies discussed above and shown in Fig. \ref{fig3:Canonical_Events_from_Catalogue}. \newline

Beginning with the 2003 event, in Fig. \ref{Fig5:Canonical_Events_case_study_2003} we show the period from the 2nd to 28th of July, which includes an event from the AA-derived catalogue associated Archetype 1 (12th to 18th July). Temperatures anomalies up to 5$^{\circ}$C are found in between 0$^{\circ}$ and 20$^{\circ}$E , corresponding to a large area of western and central Europe. Concurrently, the signature of coherent RWPs is evident in the alternating positive and negative 200hPa meridional wind anomalies and the RWP amplitude. The RWP is evident prior to the AA derived event, propagating from west to east between the 6th and 12th of July, before becoming relatively stationary between 60$^{\circ}$W and 0$^{\circ}$E. The stationary RWP, which is associated with the presence of blocking in over central-western Europe, endures for the majority of the event, and its western extent corresponds to the longitudes with the highest temperature anomalies. \newline

Similar spatio-temporal structures can be seen in both the 2010 and 2021 heatwaves (Figs. \ref{Fig6:Canonical_Events_case_study_2010} and \ref{Fig7:Canonical_Events_case_study_2021} respectively). In both events, stationary high surface temperatures are found over a constrained set of longitudes, approximately 20$^{\circ}$E -- 60$^{\circ}$E in the 2010 case, and both 115$^{\circ}$W -- 130$^{\circ}$W and 0$^{\circ}$E -- 60$^{\circ}$E in the 2021. During both events, all regions subject to extreme temperatures are under the influence of significant and relatively stationary RWPs, blocking high pressure systems, and meandering upper level winds. The the largest temperature anomalies are found to the east of the RWP envelope, that are subject to northerly upper-tropopheric wind anomalies. \newline 

During the 2010 period shown in Fig. \ref{Fig6:Canonical_Events_case_study_2010}, a stationary RWP forms just prior to the onset of the AA-identified event, becomes stationary at a longitude slightly to the east of 0$^{\circ}$, and persists at these longitudes for the duration of the event. In the 2021 case, one RWP propagates into the eastern warm region (longitudes of 0$^{\circ}$E -- 60$^{\circ}$E) from the west before becoming stationary and dissipating around the 23rd of May, while a second RWP forms on the 3rd of May over the warm anomalies found over North America (longitudes 115$^{\circ}$W -- 130$^{\circ}$W), remaining stationary beyond the event duration. \newline

\section{Discussion and Conclusion} \label{Sec:Discussion_Conclusion}

We have presented an event based analysis of Northern Hemisphere continental-scale regimes associated with extreme surface temperatures. These events are identified using archetype analysis (AA), a powerful technique that is able to approximate a high dimensional dataset as a discreet number of hyper-points located at the extremities of the data's convex hull. Our event definition defines events by seeking persistent time periods when a single archetype is strongly expressed, which is a measure of atmosphere's resemblance to a finite set of archetypal states, as a opposed to a direct measure of large deviations from the climatological average. Our approach exploits the fact that outer regions of the attractor of a dynamical system are associated with the occurrence of extreme events \citep{Lucarinietal2014JSP}. AA succeeds in robustly identifying meaningful regimes of the Northern Hemisphere mid-latitude atmosphere, which had proved difficult in a recent analysis based on Markov State Modelling \citep{Springeretal2024}, where continental patterns and teleconnections were instead robustly defined. Events identified with AA show a strong statistical link between surface temperature extremes and archetype expression, with more than 50\% of event days exceeding the 90th percentile over large areas.  \newline

We have also presented a plausible dynamical interpretation of the large-scale regimes, Archetypal regimes are linked to circum-hemispheric wave patterns that are spatially amplitude modulated. In turn, the heatwave events identified by our methodology appear to be influenced by longitudinally confined RWPs that either propagate slowly or stall over a region, resulting in anomalous heat transport and high-pressure blocking features that align spatially and temporally with the largest surface temperature anomalies. These anomalies appear to be closely related to spatially-extended heatwaves, which in some cases lead to concurring or compounding extreme events in various locations. \newline

We now discuss several implications arising from our results.

\subsection{The typicality of large-scale regimes leading to surface heatwaves}

The major finding of this work is that the large scale atmospheric regimes leading to northern hemisphere surface heatwaves are typical -- that is both recurring and relatively common, with dynamic similarity between events. All events associated with one class (i.e. an archetype) tend to resemble each other. Our results is in close agreement with the notion of dynamical typicality of extreme events based large deviation theory \citep{Galfi2021PRL,Galfi2021,LucariniEtAl2023,NoyelleEtAl2023}. The fact that these large-scale extreme regimes are common means that their statistics can be approached using empirical methods applicable to relatively large sample populations, without having to resort to statistics designed for the tails of distributions. Additionally, the event based methods used here allow for standard metrics, such as the return period, to be empirically calculated. \newline

%{\color{red}VL: Given the final comment in the penultimate paragraph of this subsection, I would lead this last comment our. Additionally, GEV is useful for assessing the probability of events even larger than those already observed (it allows for prediction in a statistical sense, see discussion in Galfi et al. Complexity 2017). So I do not think that referring to it here is helpful.} \newline
%%CC: Done  

\begin{figure}[p!]
\centerline{\includegraphics[width=6.5in, height=3.0in]{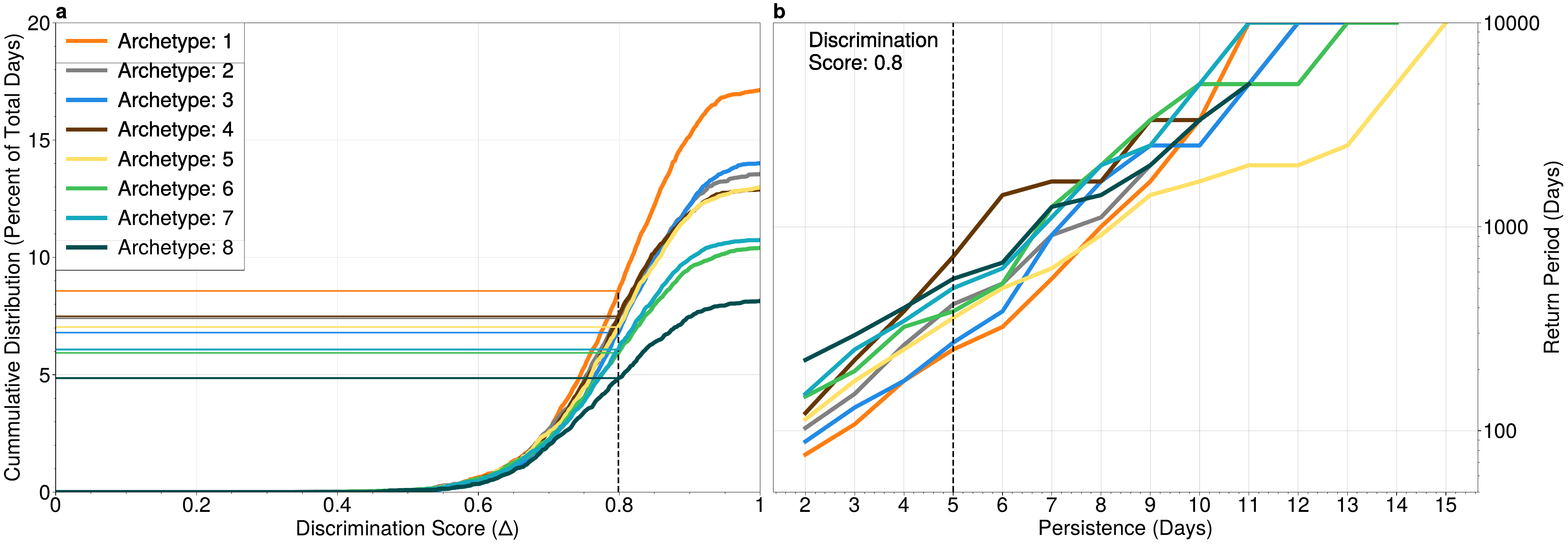}}
  \caption{\textbf{Aggregate Statistics of Extreme Regimes: } \textbf{a} Empirical cumulative distribution function (cdf) of discrimination score, clustered by `winning' archetype (i.e. that with the greatest affiliation probability), expressed as a percentage of total days in the dataset. \textbf{b} The return period in days, calculated as the number of days in the dataset divided by the number of events. }
  \label{Fig13:Aggregate Statistics}
\end{figure}

To illustrate this typicality,  we show in Fig. \ref{Fig13:Aggregate Statistics} the cumulative probability distributions for the discrimination score, and the return period of events as a function of the persistence criteria for a fixed discrimination score. The discrimination score is our measure of the similarity of the atmospheric state at any time step to a single archetypal pattern. Recalling that the discrimination score threshold of 0.8 was imposed in our definition of an event, for a given archetype, between 5 and 8.5\% of days exceed this value, resulting in a total of about 53\% of extended summer days potentially falling into an extreme state. However, the imposition of a persistence criterion dramatically reduces this number when considering events, which we see in the return periods, which we calculate by simply dividing the number of days in the dataset by the number of events. Return periods, shown inf Fig. \ref{Fig13:Aggregate Statistics}\textbf{b} are computed using persistence criteria from 2 and 16 days (the maximum detected), and a fixed discrimination score of 0.8. For a persistence criterion of 5 days (used in this study), return periods range from 150 to 800 days, depending on the archetype, giving a rough range of one single event per summer for the most common regimes, through to a one event every 4-5 extended summers for the least common. The return period grows following a power law as the persistence is increased. The relative rarity of events persisting longer than 10 days points to the dominant influence of synoptic atmospheric dynamics in driving these events.      \newline 

Archetypes are statistically and dynamically linked to extreme surface air temperatures. However, not every event in our catalogue leads to temperature extremes, and not every temperature extreme is found in our event catalogue. As we compute archetypes over the entire Northern Hemisphere, we naturally extract only the largest-scale atmospheric state, while a substantial component of an individual event is driven by regional and local processes. As such, we favour a statistical interpretation of the archetypal events: the occurrence of these relatively common regimes, expressed over an entire hemisphere, is associated with conditions that are favourable to the occurrence of extreme temperatures in certain geographical regions. 

%The results presented here lead us to use the term \textit{archetypicality}, to describe these patterns, as they are both regularly occurring and constructed from a set of representative flow configurations.  

\subsection{The role of large-scale wave modes in extreme dynamics}

Circum-hemispheric wave modes are a feature of the mid-latitude climate system \citep{bra02:jcli}, and have been implicated in surface temperature extremes. \citet{KornhuberEtAl2017} and \citet{KornhuberEtAl2019}, building off the theoretical results by \cite{PetoukhovEtAl2013}, demonstrated the importance of the wavenumber 7 patter in the development of several European heatwaves, including the 2003 event investigated here. In parallel, \cite{FragkoulidisEtAl2018} and \cite{WickerEtAl2024} provide evidence that the modulating envelope of these wave patterns, the RWP, dictates the timing, location, and intensity of any resulting extreme. \newline

Our results may provide a unification of these two related paradigms. The archetypes show hemispheric-scale meandering circulation, with wavenumbers falling typically between 4 and 6, although the energy is typically spread around a single peak. However, Fig. \ref{fig8:v200_S_composites} shows that these wave patterns are strongly modulated, with clear wave packets centered over discrete longitudes which align with the regions subject to the highest temperatures when the archetypal pattern is strongly expressed. Case studies (see Figs. \ref{Fig5:Canonical_Events_case_study_2003}, \ref{Fig6:Canonical_Events_case_study_2010}, and \ref{Fig7:Canonical_Events_case_study_2021}) show that extreme temperatures occur when upstream and downstrean of stalled RWPs. As such, our results indicate the importance of global, non-linear wave structures in the formation of surface extremes. \newline

\subsection{Limitations and scope for future work.}

There are several limitations to this work. Notably, while we note that the AA-derived events are almost always associated with extremes, not all extremes are associated with AA-derived events (see, for instance, the discussion of the August 2010 case study in Section \ref{Sec:Dynamics}). We hypothesize that this is due to the fact that the events in our catalogue are developed from the entire Northern Hemisphere, meaning that only the largest-scale atmospheric regimes are captured, and events without a global imprint are likely to be excluded. However, this point requires further investigation to demonstrate the utility of the outside-in approach for regional extremes. Additionally, our study has used only a single variable, surface air temperature, to define the large-scale regimes. However, the use of additional variables, such as geopotential height, either alone or together in a multivariate AA (described in \citet{bla22:aies}), could yield further insights into the robustness and generality of the patterns identified here. Increasing the number of archetypes in the decomposition could also lead to new event types and a greater ability to link regional events to large-scale flow states. Finally, we have almost completely neglected the temporal evolution of the archetype affiliation probabilities during the lifetime of an event. The temporal component of the affiliation time series may be instructive when considering the onset persistence and decay of regimes. \newline

Our results provide several promising avenues for further work beyond simply extending these results to the other variables (i.e. rainfall) or new regions, such as the Southern Hemisphere. For example, it is well known that in the mid-latitude atmosphere, phenomena with larger spatial scales tend to evolve more slowly and persist for longer than smaller-scale phenomena \citep{lor69:tel,Tribbia&Baumhefner2004,FranzkeEtAl2020}. Indeed, actionable extended range predictions of large scale flow states, such as those associated with the 2010 Russian heatwave, is possible in some state-of-the-art prediction systems \citep{Vitart&Robertson2018}. Our AA based approach provides a potential route to long range (i.e. beyond the usual 7-10 window of standard weather forecasts) early warning system for large-scale events. For example, extended range forecasts could be filtered through an archetypal lens to rapidly identify regions at risk of extreme events, without undue emphasis on the details, such as the precise positioning of a RWP or blocking high, which extended-range forecasting systems are unlikely to predict in an case. Such an approach could also be used to generate boosted ensembles, as described by \cite{Fisher2023}, to generate physically plausible storylines of extremes.  Additionally, Fig. \ref{fig2:Event_Lego_Plot_detrend} shows the possibility of interdecadal and decadal modulation of the presence or absence of events, which we have noted but not investigated. There is the potential for the low-frequency variability of extreme flow configurations to be linked to climate modes, such as the North Atlantic Oscillation, which could in-turn improve extended range early warning systems. \newline 

There is further potential for using this approach to assess the ability of large scale climate models to simulate both the spatial and temporal variability associated with extreme conditions. As shown by \citet{cha22:ncom} when applying AA to sea-surface temperatures in a long control run of climate model, bias resulted in spatial patterns associated with marine heatwaves being offset with respect to those obtained from satellite observations, which resulted in extreme events being favoured in the `wrong' locations. The approach taken in this work could provide useful guidance for assessing a climate model's ability to simulate the broad scale patterns associated with certain extreme events, their global teleconnections, and hence the capacity of climate projections to reliably simulate extreme states. The application of AA to climate projections, following the approach that \citet{MannEtAl2017} and \citet{MannEtAl2018},  would also allow for the assessment of changes in circulation regimes as the climate warms. \newline

\section{Appendix A: Sensitivity and Robustness of Archetypal Patterns to the Number of Archetypes}

As noted in \citet{han17:jcli}, \citet{bla22:aies}, and \citet{mon24:arxiv} the number of archetypes, known formally as the cardinality, is crucial. Unlike methods such as Principal Component Analysis that are more familiar in climate science, AA is non-orthogonal and does not preserve variance. As such, there is no guarantee that a AA decomposition is nested: that is, a decomposition with a cardinality of $p+1$ will not necessarily produce $p$ archetypes that are similar to those obtained from an decomposition with a cardinality of $p$ \citep{cut94:tec}. In practice however, archetypes frequently nest: for a decomposition using $p+1$ archetypes, the first $p$ archetypal patterns will be essentially identical to those obtained from a decomposition with cardinality $p$, and the $p$+1th pattern will add a new regime. \newline

\begin{figure}[p!]
%\centerline{\includegraphics[width=7.0in, height=7.0in]{aa_jra55_tmp_sfc_1d_dtd_NH_MJJAS_nesting_plot.C.markup.pdf}}
\centerline{\includegraphics[width=\textwidth]{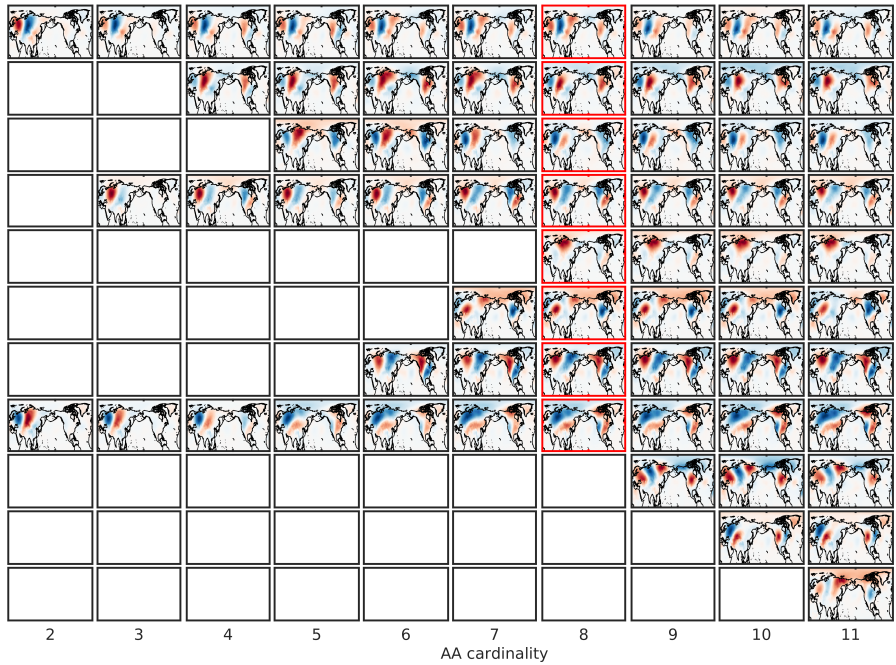}}
\caption{Matched archetypes using Pearson pattern correlation for cardinalities ranging from 2 to 11. The 10 columns by 11 rows correspond to matched archetypal patterns referenced to the AA results for cardinality of 11 (last column) across cardinalities ranging from 2 (first column) to 11 (last column). The $p=8$ column is highlighted as it corresponds to the cardinality used in the main text. The archetypes for cardinality 8 are now ordered from the most prevalent (1st row) to the least prevalent (8th row).} \label{Fig:A1_nesting}
\end{figure}

For the case of detrended, daily surface temperature anomalies, we test for nesting for cardinalities between 2 and 11. To do so, we apply a distance matching metric to the all archetypal spatial patterns for cardinalities $p=2\ldots11$, comparing them with the patterns extracted for $p=11$ (note that both forward and backward matching, as well as multiple different metrics, have been trialed). The best matching are arranged by columns in Fig \ref{Fig:A1_nesting}. In this figure, we see strong nesting up until a cardinality of 11. \newline 

Notably, the archetypal patterns associated with the 2003 western European and the 2021 western North American heatwaves appear with a cardinality of 2, while the pattern associated with the 2010 Russian heatwave appears in a cardinality of 3. However, we note that for cardinalities lower than 4, the Archetype 1 pattern is strongly weighted towards Eurasia, with minimal expression over North America. In order to capture the 2021 western North American heatwave, a cardinality of at least 4 is required. This example demonstrates the importance of choosing an appropriate cardinality for the problem at hand. \newline

At a cardinality of 8, a broad range of geographical events with good discrimination can be identified. However, up to 11 archetypes could have been employed without adding redundant information. As such, the choice of a cardinality of 8 in this case could be see as being somewhat arbitrary. We note that for cardinalities above 11, the nesting of the system begins to break down. as such, there is limited value in increasing the cardinality beyond $p=11$. 

\newpage
\bibliographystyle{apalike}
\bibliography{enso_aa}

%\begin{thebibliography}{99}

%\bibitem{1} Spiegel, M. R. (1981). Theory and problems of Advanced Calculus: Si (metric) edition. McGraw-Hill. 

%\end{thebibliography}
\end{document}